\documentstyle[12pt]{article}
\newcommand{\nc}{\newcommand}
\nc{\beq}{\begin{equation}}
\nc{\eeq}{\end{equation}}
\nc{\beqa}{\begin{eqnarray}}
\nc{\eeqa}{\end{eqnarray}}
\nc{\lra}{\leftrightarrow}

\def\muu{m_\phi}
\nc{\sss}{\scriptscriptstyle}

\newcommand{\sperp}{{\scriptscriptstyle\perp}}
\newcommand{\erf}{{\rm erf}}

\newcommand\lsim{\mathrel{\rlap{\lower4pt\hbox{\hskip1pt$\sim$}}
    \raise1pt\hbox{$<$}}}
\newcommand\gsim{\mathrel{\rlap{\lower4pt\hbox{\hskip1pt$\sim$}}
    \raise1pt\hbox{$>$}}}

\begin{document}

\input epsf.tex
\title{\vskip2cm
{\bf String-Mediated~Electroweak~Baryogenesis:\\ A Critical
Analysis}} 

\author{ J. M. Cline\footnote{e-mail:
        jcline@physics.mcgill.ca} ${}^{,a}$, 
J. R. Espinosa\footnote{e-mail: espinosa@mail.cern.ch} ${}^{,b}$, 
G. D. Moore\footnote{e-mail: guymoore@physics.mcgill.ca} ${}^{,a}$ and A.
Riotto\footnote{e-mail: riotto@nxth04.cern.ch. On leave of absence from 
Theoretical Physics
Department, University of Oxford, UK.} ${}^{,b}$
\hspace{3cm}\\
${}^a$ {\small Dept. of Physics, McGill University, 3600 University St.}\\
	{\small	Montreal, QC H3A 2T8 Canada}\\
	${}^b$  {\small CERN TH-Division, CH-1211 Geneva 23}\\
	{\small	 Switzerland}}

\maketitle
\begin{abstract}
We study the scenario of electroweak baryogenesis mediated by
nonsuperconducting cosmic strings.  This idea relies upon electroweak
symmetry being restored in a region around the core of the topological
defect so that, within this region, the rate of baryon number violation
is enhanced. We compute numerically how effectively baryon number is
violated along a cosmic string, at an epoch when the baryon number
violation rate elsewhere is negligible. We show that $B$-violation
along nonsuperconducting strings is quite inefficient.
When proper accounting is
taken of the velocity dependence of the baryon number production by
strings, it proves too small to explain the observed abundance by at
least ten orders of magnitude, whether the strings are in the friction
dominated or the scaling regime.
\end{abstract}

\vskip1cm
\leftline{}
\leftline{CERN-TH/98-306}
\leftline{October 1998}

\vskip-21cm
\rightline{}  
\rightline{CERN-TH/98-306}
\rightline{IEM-FT-182/98}
\rightline{MCGILL-98/24}
\rightline{hep-ph/9810261}

\newpage

\section{Introduction}

Considerations about the formation of light element abundances when the
temperature of the Universe was about 1 MeV lead us to conclude that the
difference between the number density of baryons and that of antibaryons
is about $10^{-10}$ if normalized to the entropy density of the Universe
\cite{nucleosynthesis}.

The theories that explain how to produce such a tiny number go
generically under the name of theories of baryogenesis, and they
represent, perhaps, the best example of the interplay between particle
physics and cosmology.  Until now, many mechanisms for the generation
of the baryon asymmetry have been proposed. Grand Unified Theories
unify the strong and the electroweak interactions and predict baryon
number violation at the tree level. They are, therefore, good
candidates for a theory of baryogenesis. There, the out-of-equilibrium
decay of superheavy particles can explain the observed baryon asymmetry
\cite{decay}.

An alternative possibility is represented by electroweak baryogenesis,
which has been the subject of intense activity in the last few years
\cite{review1,review2}. 
The Standard Model (SM) itself provides
one of the key ingredients for baryogenesis \cite{sak}:  violation of
baryon number via the anomaly ~\cite{ho}, which also constrains some
models in which the baryon asymmetry is generated at a very high energy
scale.

The underlying idea is this. If the electroweak phase
transition is of the first order, the minimum of the Higgs potential
associated with the symmetric phase is separated from the local minimum
of the broken phase by an energy barrier. At the critical temperature
$T_ c$ both phases are degenerate in energy and at later times the
broken phase becomes the global minimum of the potential. 
The universe supercools in the symmetric phase, and the phase
transition then proceeds by nucleation of critical bubbles.  As the
bubble walls separating the broken from the unbroken phase pass each
point in space, the order parameter changes rapidly, leading to a
significant departure from thermal equilibrium.  CP-violating sources
may be locally induced by the passage of the bubble wall and
baryogenesis is enhanced if the CP-violating charges can efficiently
diffuse in front of the advancing bubble wall, where anomalous
electroweak baryon number 
violating processes induced by sphaleron transitions
are unsuppressed \cite{krs}.  

In the SM, the electroweak phase transition is too weakly first order
to assure the preservation of the generated baryon asymmetry, as
perturbative and non-perturbative analyses have shown \cite{review1}.
For this reason, electroweak baryogenesis requires new physics at the
electroweak scale. Low-energy supersymmetry is a well-motivated
possibility, and there is still some room for a suitably strong phase
transition in this scenario~\cite{rev} and for the production of the
baryon asymmetry~\cite{review2}.

In this paper we will focus on a different mechanism to
implement electroweak baryogenesis, in which the third Sakharov
condition, loss of thermal equilibrium \cite{sak}, is fulfilled by the
evolution of a network of topological defects \cite{Trodden,Tomislav}.
The advantage of this idea is that it does not depend upon the order of
the electroweak phase transition, although the presence of stable
topological defects still requires new physics beyond the SM.

Topological defects are regions of trapped energy density which may be
left over after a cosmological phase transition if the topology of the
vacuum of the theory is nontrivial. In particular, cosmic strings are
solitonic solutions in spontaneously broken field theories for which
there exist non-contractible loops in the vacuum manifold
\cite{Kibblereview}.

An essential ingredient of cosmic string mediated baryogenesis is the
restoration of electroweak symmetry in a region surrounding the core of
the string.  This obviously requires some coupling of the fields
responsible for the strings, which are 
presumed to have formed at a scale above the weak scale $v$, to the
Higgs field.  For example, the strings may arise from the breaking of 
an extra gauged $U(1)'$
symmetry by the formation of a condensate of some other scalar field
$S$; if the SM Higgs is charged under $U(1)'$, electroweak
symmetry is restored in a region about the string because this reduces
Higgs field gradient energies within the string \cite{DavisPerkins}.  

After the electroweak phase transition, baryon number violation may be
efficient only in the symmetry restored region around the string, and
not outside.  If, in addition, there are CP violating interactions
between the string and the background plasma, the motion of the string
can disturb equilibrium in a way which creates a localized CP
asymmetry.  The CP asymmetry biases electroweak sphalerons to generate a
baryon asymmetry \cite{Trodden}, in a manner analogous to the physics at
a bubble wall in the conventional scenario with a first order phase
transition.  We anticipate that the asymmetry
generated by cosmic strings will be smaller because only a fraction of
space is swept out by the moving network of strings, and because the
opposite walls of the string produce asymmetries of opposite sign that
tend to cancel unless baryon number violation is very rapid inside the
string.

{}From these considerations one sees that a crucial requirement for
string mediated electroweak baryogenesis is that the baryon number violating
sphaleron transitions are fast in the regions near the strings.
Further, this must be true for some part of the epoch where baryon
number violation is very inefficient in the bulk plasma, because when
baryon number violation is at all efficient in the bulk, any asymmetry
produced by strings is subsequently erased.  Hence, the mechanism relies
on there being some temperature range in which both 
\begin{enumerate}
\item	Sphaleron configurations large enough to be unsuppressed fit inside
	the region of electroweak symmetry restoration, so the baryon
	number violation rate is unsuppressed close to string cores, and
\item	sphalerons {\it are} suppressed in the bulk plasma, so baryon
	number produced by strings is preserved.
\end{enumerate}
To see what is needed to satisfy the above requirements, let us consider
the typical length scale $R_s$ of symmetry restoration around the center
of nonsuperconducting strings:
\begin{equation} 
R_s \sim \frac{1}{\lambda^{1/2}
 v(T)} \, , 
\end{equation} 
%
%
where $\lambda$ is the
quartic Higgs coupling and $v(T)$ is the temperature dependent vacuum
expectation value of the Higgs field. On the other hand, the sphaleron
size in the unbroken phase is $R_{sph} \sim (g^2 T)^{-1}$. Conditions
$1.$ and $2.$ are then $R_{sph} \lsim R_{s}$ and $v(T)/T\gsim 1$,
respectively.  Together these imply 
\begin{equation} 
\lambda \lsim g^4 \, ,  
\end{equation} 
which indicates that one needs a Higgs boson which is
far too light and already ruled out by present LEP data which require
$\lambda \sim g^2$ \cite{bock}.  It might be possible to satisfy
this condition in extensions of the SM; however such a weak scalar
self-coupling leads to a very strong first order electroweak phase
transition, in which case baryogenesis could proceed without strings
anyway.  

It is therefore clear that a more rigorous treatment is needed to
understand whether string mediated electroweak baryogenesis can really
explain the observed baryon asymmetry. To answer this question, we have
computed either the sphaleron energy or the sphaleron rate in the
background of each sort of  nonsuperconducting strings.  For sphaleron
energies we use techniques similar to those in \cite{Baal} and for
sphaleron rates we use the techniques of \cite{slavepaper}. The case of
superconducting strings is more complicated, because it involves
determining the size of currents on such strings in a realistic
cosmological setting.  Since we are not aware of any reliable estimates
of the current carried by a superconducting string in a realistic early
universe cosmological setting we have not attempted this here and we
leave it for future investigation.

Our findings indicate that the baryon number violation along a string
is generally too slow for significant baryogenesis.  This is because
the region where electroweak symmetry is restored is never wide enough
to allow symmetric phase sphaleron-like events. At best it brings the
sphaleron rate to $1/30$ of what it would be if the core region were of
the order of the natural nonperturbative length scale $1 / \alpha_W T$;
and this only occurs in a rather
special case where the Higgs field is charged under an extra $U(1)'$,
with a charge incommensurate with the charge of the scalar field which
breaks the extra $U(1)'$.  If the charge is commensurate, or if the
Higgs field only interacts with the string generating fields via
potential terms, then the sphaleron rate is insignificant.

Moreover, if the network is in the scaling regime, the baryon number
violation occurs in too small a total volume to explain the observed
baryon asymmetry.  If the network evolution is friction dominated, the
network will be denser, but strings will move more slowly.
As we discuss in some detail here, the baryon number production is
suppressed by the second power of the string velocity, so this case also
leads to insignificant baryogenesis.
In either case we find that the generated baryon number is at least ten
orders of magnitude smaller than the observed abundance.

The paper is organized as follows: section 2 is dedicated to the
computation of the baryon number violation on the strings for different
cases; section 3 deals with the generation of CP asymmetry inside the
strings; and section 4 with the efficiency of baryogenesis by a string
network. Finally, in section 5 we present our conclusions. 

\section{Baryon number violation on cosmic strings}
\label{Bviol}

To determine how efficiently cosmic strings can generate the
baryon number asymmetry, we need to know the rate of $B$ violation by
sphalerons inside the string, at times when the Higgs field condensate 
is large enough so $B$ violation is negligibly slow outside the string.  
We characterize the efficiency of baryon number
violation on the string by the Chern-Simons number ($N_{\rm CS}$)
diffusion constant per unit string length $L$,
\begin{equation}
	\Gamma_l \equiv \lim_{t \rightarrow \infty} \frac{ 
	\langle ( N_{\rm CS}(t) - N_{\rm CS}(0) )^2 \rangle}{Lt} \, .
\end{equation}
This is related to baryon number production by the string through
\begin{equation}
\frac{1}{L}\,\frac{dN_{\rm B}}{dt} = - \frac{\Gamma_l N_F}{2} 
	\sum_i \frac{\mu_i}{T} \, ,
\end{equation}
where $N_F = 3$ is the number of families and $\mu_i$ is half the
difference in chemical potential between particle and antiparticle,
averaged over the region of symmetry restoration.  The sum runs over
each doublet which couples to $SU(2)_L$.  We will often write
$\Gamma_l$ in terms of a dimensionless quantity $\kappa_l$,
\begin{equation}
\Gamma_l = \kappa_l \alpha_w^2 T^2 \, .
\end{equation}
For a very thick string, $\kappa_l$ would numerically be approximately
$\alpha_w^2 T^2$ times the cross sectional area, although
parametrically there is actually an extra power of $\alpha_w$
\cite{ASY,HuetSon,Son,particles,Bodek_log}.  We would say that baryon
number violation is unsuppressed on the string
if $\kappa_l \sim 1$.

Since the only baryons which survive are those created after the Higgs
VEV exceeds the temperature, we will focus on temperatures such that
$\phi_0 / T \gsim 1$, where $\phi_0$ is the amplitude of the Higgs
field condensate, $\phi_0^2 = 2 \Phi^\dagger \Phi\, ({\rm broken \; phase
} )$.  We expect (and can check) that at lower temperatures, baryon
number violation 
becomes less efficient along a cosmic string.  For instance, a
cosmic string today, at temperatures $\lsim 1$MeV, will not catalyze
baryon number destruction except by extremely inefficient instanton
processes, unless the region of symmetry restoration is
very large--either $> 1 / \alpha_W T$, so sphaleron processes will be
unsuppressed inside, or $> 1 / \Lambda_{\rm EW}$, the scale at which
the electroweak coupling becomes large and instantons become
efficient.  (This scale is enormous, $1 / \lambda_{\rm EW} > 1$ meter.)

Furthermore, we will focus on the case $m_H \sim m_W$, so the scalar
self-coupling is of order the gauge coupling.  
Smaller values of $m_H$ are experimentally excluded, and larger ones 
can only decrease the efficiency of baryogenesis.
In parametric estimates we will always write $g$, meaning
either $g$, $\sqrt{g^2 + g'^2}$, or $\sqrt{\lambda}$.

Except perhaps for superconducting cosmic strings, the width of the
region of symmetry restoration on a cosmic string is of order the Higgs
field correlation length $1/ g\phi_0$, which by no coincidence is the
characteristic size of a sphaleron in the broken phase.  So it is not
{\it a priori} clear that sphaleron processes are unsuppressed on
cosmic strings.  Rather, we should check whether this is the case.  We
now endeavor to do so, case by case for various symmetry-restoring
types of strings.

\subsection{Strings which force $\Phi=0$ on the string core}
\label{thinstring}

We begin with strings which force the Higgs condensate to zero along the
core of the string, assumed much thinner than $1/g\phi_0$, but do not
influence electroweak physics outside the string core.  There are two
obvious examples of such strings.  

In the first, the standard model is extended by a complex scalar field
$S$, which is either a singlet under all gauge interactions, or
transforms under an extra $U(1)'$ symmetry for which all the standard
model particles are uncharged.  The effective potential for $S$ and the
standard model Higgs field $\Phi$ is
\begin{equation}
\label{potential}
V(S,\Phi) = \lambda_s \left( S^* S - S_0^2\right)^2 
	- \gamma ( S^* S - S_0^2) (\Phi^\dagger \Phi- \Phi_0^2) 
	+ \lambda \left(  \Phi^\dagger \Phi - \Phi_0^2 \right)^2\, ,
\end{equation}
where we have assumed that the coefficient of the interaction term
between the $S$ and Higgs fields is negative; if it is not, electroweak
symmetry will not be restored.  Here and throughout we will use the
notation in which capital letters $S$ and $\Phi$ denote complex fields,
while lower case $s$ and $\phi$ are real components in the direction of
the condensate, normalized such that $\phi_0^2 = 2 \Phi_0^2$.

Strings will form at some high temperature $T \sim \sqrt{\lambda_s} S_0
$ where the $S$ field condenses.  If the $S$ field has no gauge
interactions, the strings are global; if it transforms under a local
extra $U(1)'$ then they are local (Abelian Higgs model) strings.  In
either case, by the electroweak scale, the $S$ field carries a
condensate $S_0 \equiv \sqrt{|\langle S \rangle|^2}$.
The doublet Higgs mass parameter is $-2\lambda \Phi_0^2 + \gamma (S_0^2
- |S|^2)$.  In the bulk, where $S=S_0$, electroweak symmetry will break
because the Higgs field mass term has the usual form, $-2\lambda
\Phi_0^2$.  However, in the core of a cosmic string where the $S$ field
condensate vanishes, the Higgs field feels a large positive mass squared
$\simeq \gamma S_0^2$, 
which may be enough to force $\Phi = 0$ inside the string core.
Whether it does so or not is determined by the
competition between the Higgs field potential and gradient
energies.  The outcome of this competition depends on the values of the
quartic couplings and $S_0$, which also determine the thickness of the
string core.  We will discuss this further, in particular the
possibility that the $S$ field string is fat, in the subsection on fat
strings below.  In the present section we will restrict our attention
to the case where the $S$ field string is thin compared to the inverse
weak scale, $1/\phi_0$, and the string does successfully force $\Phi =
0$ in its core.

Another type of string which forces $\Phi = 0$ in its core was 
considered in ref.~\cite{DavisPerkins}.  The idea is that
there is a new $U(1)'$ gauge symmetry, broken by a scalar $S$ with
charge $q_s$ under $U(1)'$ but neutral under the SM gauge group.  The
SM Higgs field is assumed to also have a charge $q_\phi \neq 0$ under
$U(1)'$.  The $U(1)'$ breaks at some scale above the electroweak scale,
where the $S$ field condenses with $S_0\gg \Phi_0$.  A cosmic string will
carry a $U(1)'$ magnetic flux of $2 \pi / q_s$, and the phase of the
Higgs field gets shifted on parallel transportation around the string
by $2\pi q_\phi / q_s$.  For now we will assume that $q_\phi / q_s$ is
an integer.  If it is not an integer, the Higgs field is forced to be
small in a wider region around the string, and a $Z$ magnetic flux will
also be established; we treat this special situation in the next
section.

In the present case, the Higgs field minimizes gradient energies by
having its phase wind by $-2 \pi q_\phi / q_s$ in going around the
string, to compensate for the phase induced by the connection.  This
prevents azimuthal gradient energies in the Higgs condensate from
occurring outside the string core.  But inside, where a loop does not
enclose the complete $U(1)'$ magnetic flux, a nonzero Higgs condensate,
if it existed, would possess a large, energetically expensive azimuthal
covariant gradient, so the Higgs condensate is forced to zero.  Writing
the $U(1)'$ field strength as $B_\theta = b(r) / (q_s r)$ and the Higgs
field as 
$\Phi(r , \theta) = h(r) \Phi_0 \exp(-i\theta q_\phi / q_s)$, 
the Higgs field gradient energy per unit length of string is
\begin{equation} 
\frac{dE_{\rm gradient}}{dL} = 2 \pi
	\int r dr \left\{
	\frac{ \Phi_0^2 (1-b(r))^2 q_{\phi}^2}{ q_s^2} 
	\frac{h(r)^2}{r^2} + \Phi_0^2 h'(r)^2  \right\} \, .
\end{equation}
The very large coefficient on the first term at small $r$, where $b \ne
1$, forces $h \simeq 0$ here, but $h$ is free to rise away from the
core.  Far 
outside the cosmic string, we can redefine the Higgs field to include
the phase which compensates for the $U(1)'$ connection, and electroweak
physics then looks normal, even though 
$\Phi$ is pinned to zero inside the core of the string.

The above scenario is phenomenologically constrained.  Since the Higgs
field is charged under the extra $U(1)'$, and it couples to the
standard model fermions, they must carry $U(1)'$ charges as well .
Because the Higgs field Yukawa couplings to fermions mix the
generations, these charges must be generation-independent.  Then the
only anomaly-free charge assignment for the fermions is $q_\phi$ times
their hypercharges.  Thus the $U(1)'$ gauge field is a
standard-model-like $Z'$, which is experimentally excluded for masses
up to about $700$ GeV.  The symmetry breaking scales of $U(1)'$ and the
electroweak sector have to be separated by at least one order of
magnitude, which ensures that this string generation mechanism will
always fall into the ``thin string'' case.  (In the case of a $U(1)'$
whose generator is not proportional to that of hypercharge, anomalies
might be cancelled by adding additional heavy fermions.  The mass limits
on these other kinds of $Z'$ bosons are similar.)

For either of the two possibilities described above, the SM Higgs field
is forced to be exactly zero inside the core of a narrow string of
thickness $r_0$, and it
rises to its normal condensate value (at the given temperature)
outside.  The Higgs field profile is that which minimizes the energy
\begin{equation}
\frac{dE}{dL} = 2 \pi \int_{r_0}^{\infty} r dr \left\{ \Phi_0^2 h'^2(r) 
	+ V ( h \Phi_0 ) \right\} \, .
\end{equation}

If we assume that $dh/dr \lsim h/r$ then the first term behaves as
$dr/r$ and the second behaves as $r dr$.  Hence the potential term
dominates at large $r$ but at small $r$ it is negligible.  To see how
the profile depends on $r_0$ (the size of region where $S\ll S_0$,
which is the width of the string), assume that the role of the
potential is to force the Higgs field to reach its condensate value
within a radius of $r_1 \sim 1 / g \Phi_0$, and can be neglected for
$r<r_1$.  Then the energy is roughly
\begin{equation}
\frac{dE}{2 \pi dL} \sim \Phi_0^2 \int_{\ln r_0}^{\ln r_1} 
	\left(\frac{dh}{d(\ln r)} \right)^2 d(\ln r) \, .
\end{equation}
Each equal logarithmic interval of $r$ contributes a comparable amount
to the rise of $\Phi$.
Hence the size of $[h(r)-1]$ near $r=r_1$ will depend on $r_0$ through
\begin{equation}
h(r) - 1 \propto \frac{1}{\ln(r_1 - r_0) + O(1)} \, .
\end{equation}
The smaller the string core is, the more of the Higgs field's rise
occurs at small $r$, and the weaker the symmetry restoration at $r \sim
1 / g \Phi_0$.  But the dependence is only logarithmic in $r_0$.

At distances of order $1/g \Phi_0$ the suppression of $\Phi$ below
$\Phi_0$ is incomplete; rather than $\Phi \sim 0$, we find $\Phi /
\Phi_0 \sim 1 - 1 / \log(r/r_0)$.  But $r \sim 1 / g \Phi_0$ is the
physical scale important for setting the sphaleron rate.  This suggests
that for $r_0$ much less than  $1/ g \Phi_0$, the
energy of a sphaleron on a string will not be significantly less than
in the absence of a string.

To test this, we should determine the energy of a sphaleron in the
string background.  A sphaleron in the broken phase has a high degree
of symmetry; a sphaleron {\it Ansatz} with gauge and Higgs field profile
functions depending on radius alone \cite{Manton} describes the saddle
point unless $\lambda / g^2 \gg 1$ \cite{Yaffe}.  However the string
breaks the spherical symmetry down to cylindrical symmetry.  A
solution based on an {\it Ansatz} would need profile functions dependent on
$r$ and $z$ separately.  We prefer to find the sphaleron solution on
the lattice.  One writes down a 3D lattice Hamiltonian and seeks the
saddle point solution corresponding to the sphaleron.  We use an
$O(a^2)$-improved lattice Hamiltonian ($a$ is the lattice spacing) so
lattice spacing errors will begin at $O(a^4)$.

We make a starting guess for a sphaleron configuration by the same
technique used in \cite{slavepaper}.  Dissipative cooling will bring
this configuration close to the sphaleron, but since the sphaleron is a
saddle point and there are always round-off errors and lattice
artifacts, the configuration will miss the sphaleron and cool to the
vacuum.  An algorithm to cool toward saddle points was developed
especially for dealing with sphalerons on the lattice in \cite{Baal},
but its use becomes cumbersome for improved actions.  We use a new
algorithm for cooling to saddle points, presented in Appendix
\ref{AppendixA}.  We confirm that the candidate numerical solution for
the sphaleron really is one, by measuring its Chern-Simons number using
the technique developed in \cite{broken_nonpert}.

First, we confirm that the algorithm is working correctly by
determining the sphaleron energy in the bulk, for $\lambda / g^2
= 1/8$.  On parametric grounds the sphaleron energy can be written as
$E_{\rm sph} = 4 \pi B \phi_0 / g$, with $B$ a pure number.
We find $B = 1.815$
with good lattice spacing independence, in excellent agreement with the
result using the sphaleron {\it Ansatz}, $B = 1.82$.  

\begin{table}
\begin{center}
\begin{tabular}{|c|c|c|c|}
\hline
$a\times g\phi_0$ & $\sqrt{1.6}$ & $\sqrt{0.9}$ & $\sqrt{0.4}$ \\
\hline
$B$		  & $1.24$	 & $1.30$	& $1.37$	\\
\hline
\end{tabular}
\end{center}
\caption{Dependence of sphaleron energy $E_{\rm sph} = 4 \pi B \phi_0 /
g$, on lattice spacing for the thin core string. \label{Table1}}
\end{table}

We include the
string by pinning the Higgs field to zero along a line of lattice
sites.  The value of $B$ we obtain at three values of the lattice
spacing are given in Table \ref{Table1}.
By ``sphaleron energy'' we mean the energy of a sphaleron in
the string background, minus the energy of the string background
alone.  The weak lattice spacing dependence occurs because the
effective core size is getting narrower, since our procedure makes it
one lattice spacing wide.  It is worth commenting that the real space 
distribution of the magnetic energy associated with the sphaleron
remains fairly close to spherically symmetric; the sphaleron does not
stretch out along the string significantly.  This is also true if we
force $\Phi = 0$ by hand in some wider cylindrical region.

These results suggest that the best hope for
electroweak symmetry restoration is to have a fat string, which in turn
requires symmetry breaking of the singlet field very close to the
electroweak scale.  This case demands a more careful analysis because
the back reaction of electroweak fields on the fields generating the
string cannot be neglected, and we will return to it in a later
section.  As for the case where the Higgs field is charged under an
extra $U(1)'$, we note that none of our lattices had as narrow a core as
is actually required by the $Z'$ mass bound, so the sphaleron energy
is even higher than what we have found.

The conclusion is that $E_{\rm sph, \; with \; string} \ge 0.7 E_{\rm
sph, \; broken \; phase}$, getting closer to the broken phase value as
the string's core becomes narrower.  The relation between the sphaleron
energy and the baryon number violation rate involves zero-mode
contributions which will also be smaller for the sphaleron on the
string, since it has less symmetry.  In particular, the translational
zero modes previously gave $\int d^3 x$, bounded by the region of
consideration.  This will become $l^2 L$, with $L$ the length of the
string and $l$ given roughly by the distance the sphaleron can be
translated away from being centered on the string, at an energy cost of
$T$.  We expect, roughly, $l^{-2} \sim (g \phi_0 )^2 (E_{\rm sph, \;
broken \; phase} - E_{\rm sph, \; with \; string}) / T$.  This estimate
follows from assuming that displacing the center of the sphaleron
slightly away from the string core raises the sphaleron energy by an
amount quadratic in the displacement, and a displacement by the width
of the sphaleron, $\sim 1/ g \phi_0$, changes the energy to the broken
phase (no string) value.

If we approximate the rotational zero mode contributions as being the
same as for a sphaleron in the broken phase, then we may take the
values determined in \cite{Carson}.  This gives $\kappa_l \sim
\exp(-12)$ for $v/T = 1$.  In fact the rate is smaller because the
rotational symmetry is also broken, so the rotational zero modes will
be partially removed.  We made a preliminary calculation using the
nonperturbative techniques of \cite{broken_nonpert}; the result was
$\kappa_l \sim 10^{-6 \pm 0.5}$, which is uninterestingly small, so we
did not refine the calculation further.

We conclude that if a string only influences electroweak fields in
its core, and the core width is less than the weak scale $m_W^{-1}$,
then it fails completely to allow efficient baryon number violation.

\subsection{Strings of non-integer magnetic flux}

In the previous subsection we saw that if the Higgs field transforms
under an extra $U(1)'$ with charge $q_\phi$, and a string carries a
$U(1)'$ magnetic flux of $2 \pi / q_s$ with $q_\phi / q_s$ an integer,
then the Higgs field is pinned to zero in the core of the string.  The
case where $q_\phi / q_s$ is not an integer is much more interesting,
and has been explored by Davis and Perkins \cite{DavisPerkins}.
Outside of the string, if there are no other gauge field condensates,
the Higgs field phase changes by $2 \pi q_\phi / q_s$ on parallel
transport around the string, using the gauge field as the connection.  If
the Higgs field condensate has a phase 
which changes by $-2 \pi n$ around the string, so that $\Phi(r,\theta)
= \Phi_0 h(r) \exp(- n i \theta)$, and if $n\ne q_\phi/q_s$, as will be
the case if $q_\phi/q_s$ is not an integer, then there will still be a
gradient in the azimuthal direction, leading to a gradient energy of
\begin{equation}
2 \pi \Phi_0^2 \int r dr \left\{ h'^2 + 
	( n - q_\phi / q_s )^2 \frac{h^2}{r^2} \right\} \, .
\end{equation}
Since $h(r)$ must eventually reach its asymptotic value of 1 to avoid an
extensive 
potential energy cost, this leads to a logarithmically divergent gradient
energy, the logarithm arising from $\int (1/r^2) r dr$.  
This can only be removed by having a
nonzero $Z$ boson field, with a net $Z$ field magnetic flux of $2 \pi
( n - q_\Phi / q_s ) (2/g)$, to compensate.  Since the total $Z$
magnetic flux is fixed by the requirement that the Higgs gradient energy
vanishes at large distances, $\int B_Z d^2 x$ is fixed.  But the
energy in the $Z$ magnetic field is 
$\int B_Z^2 d^2 x$, which is minimized (for fixed $\int B_Z d^2 x$)
by spreading $B_Z$ out over a large region of space. This
leads to a larger region of symmetry restoration than in the
previous section, and presumably a smaller sphaleron energy.

The string is widest 
when $q_\phi / q_s = 0.5 \; {\rm modulo} \; 1$.  For this
case, writing the $Z$ field as 
\begin{equation}
Z_{\theta} = \frac{z(r)}{ g r } \, ,
\end{equation}
(approximating that the weak hypercharge gauge coupling $g'$ is zero 
henceforth), the energy per unit length of string is
\begin{equation}
\frac{dE}{dl} = 2 \pi \int_0^{\infty} r dr \left\{
	\frac{1}{2 g^2} \frac{z'^2}{r^2} + \Phi^2_0 h'^2
	+ \frac{(z-1)^2 \Phi_0^2 h^2}{4 r^2} + 
	\lambda \Phi_0^4 (h^2 - 1)^2 \right\} \, .
\end{equation}
The solution is independent of the core radius of the string
provided that it is small enough.  We have therefore taken the 
$r_0 \to 0$ limit in this equation.

\begin{figure}[t]
\centerline{\epsfxsize=6in\epsfbox{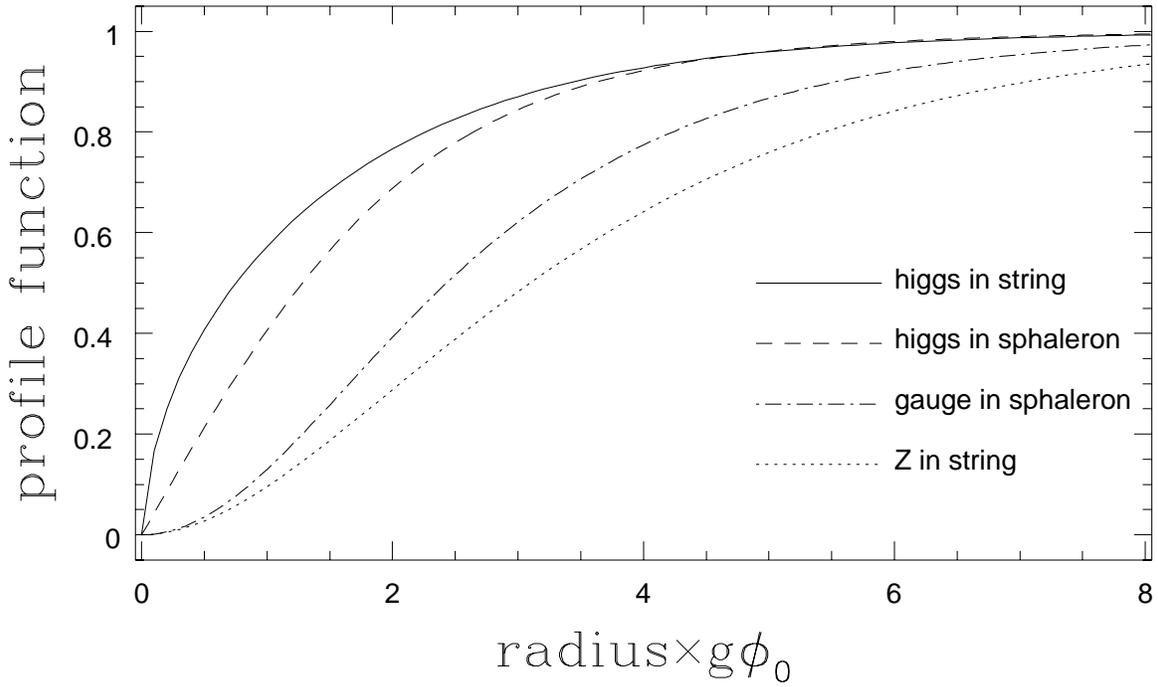}}
\caption{\label{profiles} Higgs and gauge field profile functions for
the string and the sphaleron.  The Higgs field rises faster in the
string, but the $Z$ field rises more slowly than the gauge field of the
sphaleron.  In each case, $\lambda = g^2 / 8$.}
\end{figure}

We can solve for the profile functions $z(r)$ and $h(r)$ which minimize
this energy, and compare them to the gauge and Higgs profile functions
of a sphaleron; see Figure \ref{profiles}.  The figure is slightly
misleading for two reasons:  first, the string profiles would rise
somewhat faster had we not made the approximation $g' = 0$; and second,
the string profiles are for the radial direction in cylindrical
coordinates, while the sphaleron profiles are for the radial direction
in spherical coordinates.  Hence more of the sphaleron lies within the
string than the profile functions suggest.  The string and the
sphaleron are about the same size, and we should expect an order unity
shift in the sphaleron energy in the presence of the string.  The
situation here is much more favorable than in the last subsection.

However, the above argument is too naive in the present situation.
Unlike the case in the last subsection, where the effect of the string
could be treated as just a modification of the Higgs potential along
the string core (shifting the mass up enough to pin $\Phi = 0$), here
the string background already contains large gauge magnetic fields and
azimuthal Higgs field gradients, which will influence the fields of the
sphaleron configuration, potentially changing the sphaleron
energy substantially.  Directly measuring the sphaleron energy shows
that in fact it is significantly lowered.

When the sphaleron energy becomes small (of order the temperature), the
perturbative computation of $\Gamma_l$ by the saddle point method
becomes less reliable, and we are better off using a fully
nonperturbative technique.  If the rate is not too small, we can watch
sphalerons occur during a real time evolution by tracking Chern-Simons
number topologically \cite{slavepaper}.  If the rate is so small that
no sphalerons will occur during a reasonable amount of Hamiltonian
time, we can use a new technique based on finding the sphaleron free
energy nonperturbatively on the lattice \cite{broken_nonpert}.  It
turns out in our case that for parameters which give a Higgs condensate
of $\phi_0 / T = 1$, the sphaleron rate is large enough on the string
that the real time approach can be applied.

We work on a lattice of spacing $a = 2 / ( 5 g^2 T)$, using
($O(a)$-improved \cite{Oapaper}) scalar self-coupling $\lambda =
0.125 g^2$ and a value of the Higgs mass squared parameter which gives
$v/T = 1$ (technically, $\phi^2 = 2.5 g^2 T^2$).  We set the Weinberg
angle $\Theta_W = 0$, that is, we ignore weak hypercharge, so that we 
only have to deal with one $U(1)$ gauge field.  The volume is a 3D $L_1
\times L_1 \times L_2$ torus.  We use $L_1 = L_2 = 12.8 / g^2 T$ 
(a $32^3$ lattice) and $L_2 = 14.4 / g^2 T$, $L_1 = 16 / g^2 T$ (a $36
\times 40^2$ lattice).  Both volumes are much larger than it takes to
get continuum-like behavior in sphaleron rates \cite{AmbKras}, but we
need to check whether the string's presence changes this picture.

We fix a $U(1)'$ background field appropriate for a half unit flux
along a vertical line at $(L/2 , L/2 , z)$.  For topological
reasons the integral of magnetic flux across the cross section of the
box must equal zero, so there must be a return flux somewhere; we put
it all on the plane $x = 0$ (which is identified with $x = L$) or the
plane $y=0$ (identified with $y = L$) in such a way that the connection
is undisturbed from the string solution except along these planes.  Using
two box sizes allows us to check that the return flux is not affecting the
results.  Besides this fixed U(1) background, the approach is the same
as used in \cite{slavepaper}.  In particular we do not introduce extra
degrees of freedom to generate hard thermal loop effects beyond those
induced by UV lattice modes, although the technology exists
\cite{particles}.  This makes our results an overestimate, by of order
a factor of 2.

We find diffusion constants of $\kappa_l = 0.051 \pm 0.012$ in the smaller
box, and $\kappa_l = 0.073 \pm 0.026$ in the larger box, which agree
within the statistical errors.  Baryon number violation is about 30 times
less efficient than it would be if the diameter of the symmetry restored
region were $1 / \alpha_W T$, where the usual symmetric phase rate would
be recovered \cite{slavepaper}.  The sphaleron rate on the string is
orders of magnitude larger than the rate in the broken phase, so the
string is strongly enhancing baryon number violation; however, the rate is
not quite high enough to say that sphalerons are fully unsuppressed along
the string.

We also find that $\kappa'$ is strongly dependent on $\phi_0$, and hence
on the temperature.  Going to $\phi_0 / T = 1.4$, it falls to a value of
$\kappa' = 0.028 \pm 0.008$.  Since $\phi(T) \simeq \phi(T=0) (1 - T^2 /
T_c^2)$ except very close to the phase transition temperature, and since
$T_c \simeq 100$ GeV for a realistic scalar self-coupling and top quark
mass, changing $\phi_0/T$ from $1.0$ to $1.4$ corresponds to a decrease in
temperature of approximately $8 \%$.  Thus $\kappa'$ is quite temperature
sensitive at $\phi_0 / T \simeq 1$.

\subsection{Fat strings}
\label{fatstring}

Next we return to the first example of subsection \ref{thinstring}, and
investigate the possibility that the resulting string is ``fat,'' with the
Higgs field VEV suppressed in a large region surrounding the string.  We
would like to find values of the quartic couplings and the singlet VEV
$S_0$ defining the potential (\ref{potential}) which give the largest rate
of sphaleron processes inside the largest possible strings.  This requires
that the doublet VEV $\phi(r)$ be close to zero near $r=0$, the center of
the string, and throughout a region of space large enough to allow
unsuppressed baryon number violation. 

Before examining this model's suitability for baryogenesis, it may be
worth pointing out that it cannot be obtained from supersymmetry, because
an $F$-term contribution to the potential would give a positive
$|S|^2|\Phi|^2$ coupling, whereas we need it to be negative.  A $D$-term
would give a negative coupling if the $\Phi$ and $S$ fields had opposite
charges under a $U(1)$ gauge symmetry, but this situation is not
compatible with having a fat string, for the following reasons.  Certainly
we should not identify the $U(1)$ with weak hypercharge, since the $S$
field would break it at a high scale.  On the other hand, if the standard
model Higgs field $\Phi$ transforms under a new $U(1)'$, the mass $m_{Z'}$
of the corresponding $Z'$ boson must be greater than 800 GeV, as discussed
in section 2.1; however the string width is of order $m_{Z'}^{-1}$, and so
the string will be narrow, not fat, such as we wish to construct here.

Also, if the $U(1)'$ symmetry associated with $S \rightarrow S e^{i
\theta}$ is gauged, then for fixed $S_0$ and $\lambda_s$, the resulting
string is always thinner than if the symmetry is global.  This is
particularly true if the new gauge coupling $g'$ satisfies $g'^2 \gg
\lambda_s$, as it must if $\lambda_s$ is small and $g'$ is to unify with
the standard model gauge couplings at some high scale.  As we will see,
the case of small $\lambda_s$ is the most promising.  We also observe that
the global string's fields approach their asymptotic values at $r=\infty$
slower than those of a gauged string.  The scalar condensates have
power-law behavior in global strings, $\phi(r)\sim \phi(\infty) - a
r^{-b}$, but exponential behavior in local strings, $\phi(r)\sim
\phi(\infty) - a e^{-b r}$.  Therefore global strings also have a better
chance, qualitatively, of having a large region of symmetry restoration. 
For these reasons, we will consider only global strings in the present
section.

While searching for the parameters favorable to baryogenesis, we must keep
in mind a number of constraints:
\begin{enumerate} 
\item The vacuum must be stable,
implying that
\begin{equation}
		4 \lambda \lambda_s > \gamma^2 \, .
\end{equation}
\item The mass of the lightest Higgs boson must be above the experimental
limit.  While this limit depends on the relative admixtures of the
$\Phi$ and $S$ fields in the light eigenstate, it is safe to say that
$m_h > 75 - 80$ GeV.  Solving for the mass eigenvalues gives the
constraint
\beqa
\label{higgsmass}
	\hat m_h^2 &\equiv& m^2_h/\Phi_0^2 \nonumber\\ &=&2 y_0 z
	\left( x/y_0 + y_0/x - \sqrt{(x/y_0-y_0/x)^2 + \hat\gamma^2}
	\right) > 0.2\, ,
\eeqa
where we have defined the useful combinations
\begin{eqnarray}
\label{parameters}
	x &=& \sqrt{\lambda/\lambda_s}\nonumber\\
	y &=& (S_0/\Phi_0)|_{T=T_c};\qquad 
	y_0 = (S_0/\Phi_0)|_{T=0} \nonumber\\
	z &=& \sqrt{\lambda\lambda_s}\nonumber\\
	\hat\gamma &=& \gamma/z,
\end{eqnarray}
and $\Phi_0 = 246/\sqrt{2}$ GeV $= 174$ GeV at $T=0$, but it has a smaller
value, $\sqrt{2}\Phi_0\ge T\sim 100$ GeV at the temperature where the
bulk sphaleron rate first becomes negligible.


\hspace{0.2in} Really the condition should be stronger still; we should
also demand that 
either the lighter vacuum mass eigenstate must be almost
purely $\Phi$, or both eigenstates must be fairly light.  Otherwise 
electroweak radiative corrections could be generated that would 
contradict precision tests.  (Another way of evading precision electroweak
test bounds would be to make the heavier mass eigenstate almost pure
Higgs field, with a mass of $\sim 100$ GeV;
but this turns out to require parameters that
do not lead to electroweak symmetry restoration in the string core.) 
However, in what follows we will be able to rule out string-mediated
baryogenesis without bothering to enforce this condition.

\item We require couplings to be perturbatively small, as otherwise
the theory receives large radiative corrections, and actually cannot
be defined without a UV regulator close to the energy scale of interest.
Our condition (perhaps too generous) is
\begin{equation}
	\lambda,\ \lambda_s,\ \gamma < 1 \ .
\label{pert_control}
\end{equation}
\item Finally, a less obvious but crucial requirement is that the $U(1)'$
symmetry broken by the VEV of $S$ be restored at high temperatures.
Otherwise the phase transition that should have given rise to the strings
will never have occurred.  This means that the thermal correction to
the $S$ field mass must be positive: $\delta m_s^2 = (2\lambda_s -
\gamma) T^2/6 > 0$, and hence
\begin{equation}
\label{symrest}
	\gamma < 2\lambda_s\ .
\end{equation}
This condition might be evaded by adding extra particles to the theory
which would enhance the thermal mass of the $S$, about which we will 
say more below.
\end{enumerate}

The search of parameter space can be simplified by noticing that 
a rescaling of the string's radial variable $r$ by $r\to 
r/z^{1/2}$ transforms the Hamiltonian density according to
\begin{equation}
	H(r; \lambda, \lambda_s, \gamma) \to
	z H(r/z^{1/2}, x, 1/x,\hat\gamma).
\end{equation}
This implies that the solutions for the fields around a string can be
obtained from the case where $z = 1$ by simply stretching
the spatial size by the factor $1/z^{1/2}$.  The other
important measure of symmetry restoration, the smallness of $h(r)$ at
$r=0$, is unaffected by this transformation.  So it is sufficient to vary
just the three parameters $x$, $y$ and $\hat\gamma$ of eq.~(\ref{parameters}) 
to generate all possible solutions. 

The most favorable case for baryogenesis occurs when $\hat\gamma$ is
pushed close to its upper limit of 2, coming from vacuum stability. 
Naturally one needs for $\gamma$ to be large since in the opposite limit
of $\gamma=0$ there is no coupling between the $S$ and $\Phi$ fields,
hence no possibility for electroweak symmetry restoration in the strings. 
Just at the vacuum stability limit the potential develops a flat
direction, hence a massless Higgs boson.  The experimental limit on
the Higgs boson mass, together with Eq. (\ref{pert_control}), prevent
one from taking $\hat\gamma$ higher than about 1.8.

To make $\phi(0)$ small, one also needs to have $S_0 \gg \Phi_0$ or
$y\gg 1$.  Otherwise the symmetry is only weakly or not at all restored
in the core of the string.  On the other hand, we need a wide string
core to avoid the thin string case, which already proved
unsuccessful in subsection \ref{thinstring}.  This requires that
$\lambda_s S_0^2$ not be too large, hence $\lambda_s$ must be small. 
To summarize, the most favorable situation for baryogenesis is when
\beq
\label{tuning}
	x \sim y \gg 1;\qquad \hat\gamma \lsim 2 .
\eeq

\begin{figure}[!b]
\centerline{\epsfxsize=6in\epsfbox{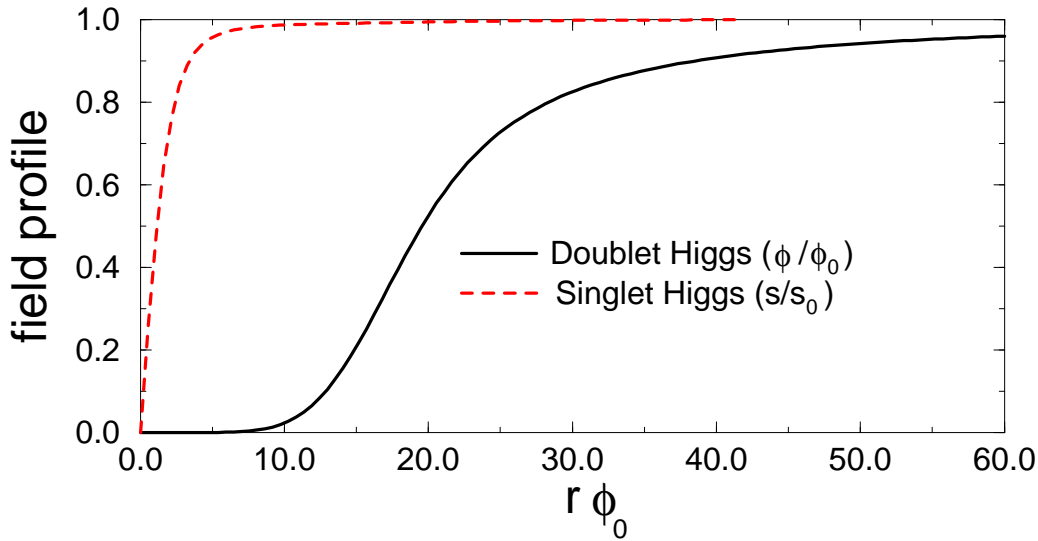}}
\caption{\label{more_profiles} Doublet and singlet Higgs field profiles
around a string, for the parameters $\lambda = 0.35$, $\lambda_s =
8.8\times 10^{-5}$, $\gamma = 9.4\times 10^{-3}$, and $S_0 = 63
\phi_0$, where $\Phi_0 = 174$ GeV.  These parameters do not allow
thermal symmetry restoration of the $S$ field.} \end{figure}

The field profiles in such a case which is favorable for baryogenesis
are shown in figure \ref{more_profiles}.  There we graph $\Phi(r)$ and
$S(r)$ versus $r$ for the couplings $\lambda = 0.35$, $\lambda_s =
8.8\times 10^{-5}$, $\gamma = 9.4\times 10^{-3}$, and $S_0 = 63 \phi_0$
(corresponding to $x = 31$, $y= 63$, $z=5.5\times 10^{3}$ and
$\hat\gamma = 1.7$).  The overall scale of the couplings is chosen to
saturate the experimental bound on the Higgs mass.  One notices that
 the region
of symmetry restoration is very wide compared to the weak scale
$\Phi_0^{-1} \cong 10^{-3}$ fm.  In this example, the half-width
$r_{1/2}$, where the Higgs field reaches a value of
$(\Phi(0)+\Phi_0)/2$, is nearly 20 times $\Phi_0^{-1}$.  To get such a
wide $\Phi(r)$ profile, it is essential that we use global rather than
gauged strings.  No radial field excitation in the broken phase is
light,\footnote{There is, however, an axial excitation--the
axion--which creates cosmological problems for these models when the
symmetry-breaking scale is low.}\
and if the $S_0$ field approached its asymptotic value exponentially,
the region of symmetry restoration could not be wider than the largest
inverse mass.

\begin{figure}[ht]
\centerline{\epsfxsize=6in\epsfbox{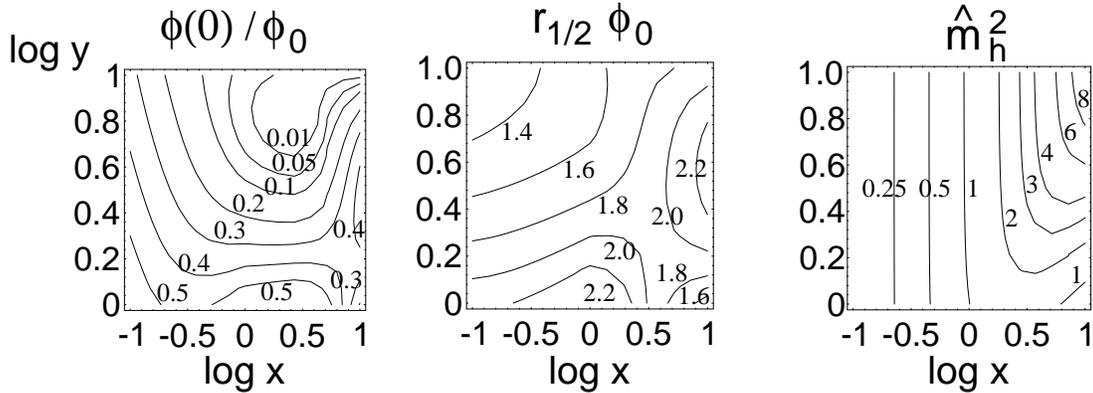}}
\caption{\label{contours1} 
(a,b) Dependence of the Higgs field central value and half-width,
$\phi(0)$, $r_{1/2}$ respectively, on the potential parameters $x =
\sqrt{\lambda/\lambda_s}$ and $y = S_0/\Phi_0$, for $\hat\gamma = 1.7$
and $z=1$. The logarithm is base 10.  (c) Contours of the dimensionless
Higgs mass $\hat m_h^2 = m_h^2/\Phi_0^2$ for the same parameters.  }
\end{figure}

Figure \ref{contours1} shows how $\Phi(0)$, $r_{1/2}$ and $\hat m_h^2$
depend on $x$ and $y$ for the choices $\hat\gamma = 1.7$ (same as in
figure 2) and $z = 1$.  The crucial observations are that $\Phi(0)$ tends
rapidly toward zero (maximum symmetry restoration in the string) as
$S_0/\Phi_0$ gets large, and the Higgs mass increases most rapidly along
the direction of increasing $x\sim y$, while the half-width varies
relatively slowly.  To increase $r_{1/2}$ dramatically, as in figure
\ref{more_profiles}, one should go to large values of $\hat m_h^2$ and
$x\sim y$, and then rescale all the couplings by $z = 0.2/\hat m_h^2$, so
as to decrease the Higgs mass to the point of saturating the experimental
bound. This increases $r_{1/2}$ by the factor $z^{-1/2}$ without changing
$\Phi(0)$.

However we have not yet taken into account the symmetry restoration bound,
Eq. (\ref{symrest}), which does not allow one to make $\lambda_s$ much
smaller than $\gamma$.  If we saturate this bound, we find that
$\hat\gamma = 2/x$, forcing us to give up at least one of the two
baryogenesis-favoring conditions, either that $\hat\gamma\sim 2$ or $x\gg
1$.  If we give up the latter condition, so that $x\sim 1$ and $\log x\sim
0$, figure 3 shows that $m_h$ does not grow with $y$, giving no leverage
to rescale $z$ and hence to widen the Higgs field profile.  Indeed,
eq.~(\ref{higgsmass}) shows that in the fixed-$x$, large-$y$ limit, $\hat
m_h^2 = xz(2-\hat\gamma^2)$.  Thus the smallest we can make $z$ is
$0.2/(2-\hat\gamma^2)$, which is of order unity or greater if $\hat\gamma$
is close to 2.  The region of symmetry restoration is of order the weak
scale in this case, which is too small to comfortably contain a sphaleron. 
In fact, the large-$y$, $x \sim 1$ case gives the thin strings we have
already considered above.  And if we choose $x \sim 1$ and $y \sim 1$,
then $\Phi(0)$ is not close to zero; the Higgs field does not lose its
condensate even in the core of the string. 

If, on the other hand, we try to keep $x$ large while decreasing $\gamma$,
the electroweak symmetry restoring effects of the string rapidly diminish,
as shown in figure 4, which is the same as figure 3 but for the smaller
coupling $\hat\gamma = 0.1$.  As the figure shows, it takes a very large
value of $y$ ($S_0$) to suppress $\Phi(0)$ in this case;  however
$r_{1/2}$ decreases with $y$ for fixed $x$, while $m^2_h$ reaches an
asymptotic value.  Taking for instance $x=20$, the largest value allowed
by the symmetry restoration bound at this value of $\gamma$, we get a
string of radius $r_{1/2}< 3.5/\phi_0$ and $\phi(0) = 0.1 \phi_0$ when
$y=100$. Both quantities decrease monotonically with $y$, which must be
large in order to suppress the Higgs field in the string core, but the
radius over which it is suppressed shrinks at large $y$.  For this value,
$x = 20$, $y = 100$, and $\gamma = 2 \lambda_s$, we find a sphaleron
energy $0.492$ times the broken phase value.  The string cuts the
sphaleron energy in half, so the sphaleron rate will be substantially
faster than the broken phase rate, but still much slower than the
symmetric phase rate, because of the Boltzmann suppression factor
$e^{-E_{\rm sph}/T}$.  The situation does not improve if we vary $y$
in either direction. 

\begin{figure}[!t]
\centerline{\epsfxsize=6in\epsfbox{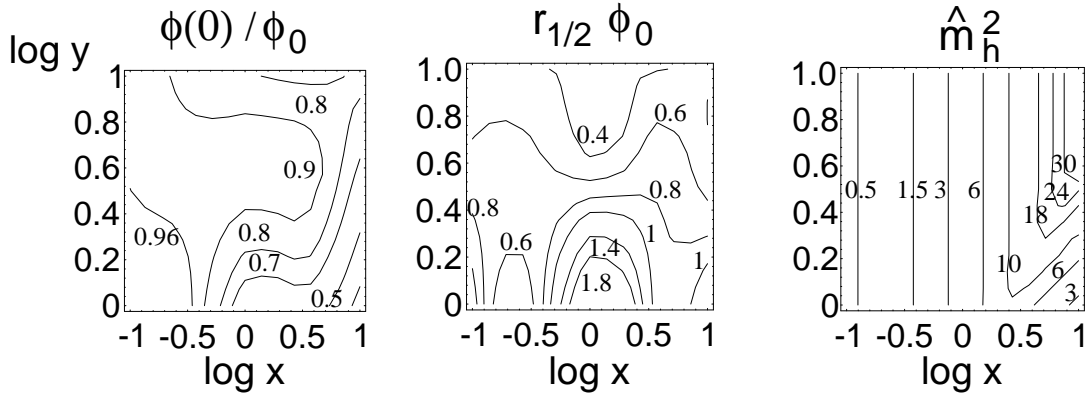}}
\caption{\label{contours2} 
Same as figure 3 but with $\hat\gamma = 0.1$  }
\end{figure}

It is possible to contrive to evade the symmetry restoration bound,
eq.~(\ref{symrest}).  One could add another scalar field $f$ with a positive
quartic coupling $\gamma' |f|^2 S^* S$, whose sole purpose is to increase the
thermal mass of the $S$ field.  The constraint is then relaxed to $\gamma <
2\lambda_s + \gamma'$ if $f$ is complex and $\gamma < 2 \lambda_s + \gamma' /
2$ if $f$ is real.  But adding $f$ introduces new problems: $\lambda_s$
receives radiative corrections of order $(\gamma')^2$, which requires
fine-tuning of the ultraviolet couplings to preserve the hierarchy $\lambda_s
\ll |\gamma|$.  Furthermore we must either tune $m_f<1$ KeV, or give the $f$
particles a direct decay channel by adding a cubic
term of form $f|\Phi|^2$ and making $m_f > 4m_b$; otherwise they leave
an excess relic abundance which overcloses the universe.  

It appears possible, then, to allow for ``fat'' strings with large regions
of symmetry restoration, but it requires a highly-tuned hierarchy of
couplings in a very artificial model.  Aside from this possibility, ``fat''
global strings do not appear to provide an effective means of restoring
electroweak symmetry in a suitably large string core to allow efficient
baryon number violation. 

To conclude this section, there are two nonsuperconducting string models
which allow significant baryon number violation along strings.  The case
where an extra $U(1)'$ is broken by a complex scalar works if the SM Higgs
field has a charge $q_\phi$ under the $U(1)'$ which is incommensurate with
the charge $q_s$ of the scalar responsible for breaking the $U(1)'$
symmetry.  The baryon number violation along such a string is much faster
than in the broken phase, but is not completely unsuppressed.  It is also
possible to violate baryon number efficiently in a global string model if
the coupling between scalar and Higgs is negative and if there is a large
hierarchy of scales between the Higgs self-coupling, the Higgs-singlet
coupling, and the singlet self-coupling.  However it is also necessary to
add a real scalar with cubic couplings and to tune them in a way
which is unstable to radiative corrections; thus the model appears to be
rather unnatural.

\section{Generation of CP asymmetry inside the string}

As we saw at the beginning of the last section, the baryon number
production is proportional to the product of the sphaleron rate along the
string and a CP-odd sum of particle chemical potentials, averaged over the
volume in which the sphalerons occur along the string.  So besides the
sphaleron rate, discussed in the last section, we also need to know how
the string biases particle populations, so as to create an asymmetry
between particles and their CP conjugates inside the string.
	
Chemical equilibrium together with CPT ensures that CP-odd combinations of
chemical potentials must vanish, so the string only generates such
particle distributions if it drives the plasma out of equilibrium, which
occurs if the string is moving relative to the plasma.  The main
goal of this section is to show that, if the string moves slowly, any CP
violating chemical potential it generates depends quadratically on the
string velocity $v_s$.  The fact that it goes as $v_s^2$ and not $v_s$
is easy to understand; a string has no natural ``front''
and ``back'' sides, so there is no distinction between it moving forward
or backward; hence the chemical potentials averaged over the string volume
must be even functions of $v_s$.  Assuming only that the chemical
potential on the string has analytic behavior in $v_s$, we get that $\mu
\sim v_s^2$ for small $v_s$.  It remains to check whether this picture
is really right, and to see how big the coefficient is.

We will treat this problem in three different ways.  First, we will
consider an unrealistic model where all parameters are pushed to the
extreme values that give the most favorable outcome for baryogenesis. 
This will be used in the following sections to make a robust argument
against the viability of string-mediated baryogenesis.  We will then make
more realistic estimates of the CP-violating chemical potentials,
$\mu_{CP}$, using a purely quantum mechanical treatment for the
particle-string interaction, as well as a semiclassical one.  It will be
seen that both of these give smaller predictions for the size of the CP
violating effects than does the toy model.  A crucial aspect of these
results is the dependence of $\mu_{CP}$ on $v_s$, the string velocity.
In particular, for small $v_s$ we will show that $\mu_{CP}
\propto v_s^2$.



\subsection{Model-independent upper limit on ${\mathbf \mu_{CP}}$}

CP violation at the string wall comes about when the probability for
particles to reflect from the wall differs from that of antiparticles. 
The most extreme situation imaginable for maximizing this effect would
occur if the string walls totally reflect particles, and
perfectly transmit the corresponding antiparticles.  We will make this
radical assumption now, in order to get a robust upper limit on
$\mu_{CP}$, hereafter simply denoted by $\mu$.  Let us now consider how
this affects the particle distributions inside the string, where baryon
number violation can occur.  Since the antiparticles do not see the
wall, they are continually refreshed by the equilibrated population
outside of the string, so their
distribution functions remain thermal.  We thus have
\beq
	f_- = {1 \over e^{E/T}\pm 1} = f_f \, ,
\eeq  
where $f_f$ denotes the equilibrium Fermi population function.

The particles, on the other hand, have in addition the effects of the
reflections, which in the rest frame of the plasma causes their
distribution functions to be boosted by the string velocity (taken to
be in the $x$ direction).  It is also possible that they acquire a
chemical potential, but other departures from equilibrium are damped by
collisions.  Assuming only that collisions with plasma particles are
less frequent than collisions with the walls of the symmetry restored
region, which is reasonable for the relatively narrow strings considered
in this paper, we get
\beq 
	f_+ = {1 \over e^{\gamma_v(E-v_s p_x +\mu)/T} \pm 1};\qquad
	\gamma_v = (1-v_s^2)^{-1/2} \, .
\eeq
For definiteness we will henceforth assume the particles are fermions, and
denote them by $F$, since only fermions enter in the anomaly equation
and bias baryon number violating processes.

The generation of the chemical potential $\mu$ is due to annihilation
($F\bar F \to GG$) and pair-production processes, ($GG\to F\bar F$),
involving gauge bosons $G$, for example.
The size of the equilibrium value of $\mu$ which is generated can be
deduced from the Boltzmann equations for the $F$ and $\bar F$
distributions: it is that value of $\mu$ which causes the production
rate of particles via $GG\to F\bar F$ and the destruction rate 
from $F\bar F \to GG$ to cancel each
other.  This is the right condition because particles inside the string
can never get out; if the production and destruction rates did not
balance, the particle number would adjust until it did.  Let $f_b = 
(e^{E/T}-1)^{-1}$ denote the thermal distribution functions for the 
gauge bosons, ${\cal M}$ the matrix element for annihilation or 
pair production, and $d\Pi \equiv \delta(p_1 + p_2 - p_3 - p_4)
\prod_{i=1}^4 d^{\,3}p_i/(16\pi^3 E_i)$ the phase
space measure.  The condition that the collision term 
vanishes is
\beqa
	\int d\Pi
	|{\cal M}|^2 \left[ f_+(p_1)f_-(p_2)(1+f_b(p_3))(1+f_b(p_4))
	\right. \nonumber\\
	\left. - f_b(p_3)f_b(p_3)(1-f_+(p_1))(1-f_-(p_2))\right] = 0.
\eeqa
Expanding $f_\pm$ in powers of $\mu$ and $v_s$, and using that $E_1 +
E_2 = E_3 + E_4$, we find after menial algebra that the term in brackets
involving population functions is
\beq
\left[ f_f(p_1) f_f(p_2) (1+f_b(p_3)) (1+f_b(p_4)) \right]
\left( \frac{\mu}{T} + v_s \frac{p_{1z}}{T} - \frac{v_s^2}{2}
	\frac{E^2}{T^2} + \frac{v_s^2}{2} \frac{p_{1z}^2}{T^2} 
	(1 - 2 f_f(p_1)) \right) \, ,
\eeq
plus corrections of order $\mu^2$, $v_s^3$, or $\mu v_s$.
Since the integration measure contains an average over angles, $p_z$
averages to zero and $p_z^2$ averages to $(1/3)p^2$.
It is because $p_z$ averages to
zero that no $v_s$ term appears in the expression for $\mu$.  It is also
clear that for all higher order processes, the term proportional to $v_s$
will involve an integration over a vector quantity and will
vanish on angular integration; so the $v_s^2$ dependence of $\mu$
is not an accident of expanding to leading parametric order.

Now, defining
\beq
\langle X \rangle_{\sss\cal M} \equiv
 {\int d\Pi |{\cal M}|^2 f_{f}(p_1)f_{f}(p_2)(1+f_{b}(p_3))
	(1+f_{b}(p_4)) X \over 
\int d\Pi |{\cal M}|^2 f_{f}(p_1)f_{f}(p_2)(1+f_{b}(p_3))
	(1+f_{b}(p_4))},
\eeq
then the result for the chemical potential can be simply expressed as
\beq
	\mu = {v_s^2\over 2}\left(\langle E_2 \rangle_{\sss\cal M}
	- {1\over 3T}\left\langle  |\vec p_2|^2 
	{e^{E_2/T} - 1\over e^{E_2/T} + 1}
 	\right\rangle_{\!\!\sss\cal M} \right)
\eeq
Notice that the result is independent of the overall strength of the
interaction, ${\cal M}$.  This will be true so long as the interaction
is faster than the Hubble expansion rate, so that it remains in
equilibrium.  The answer does involve the energy dependence of ${\cal
M}$, but the fact that $\mu \propto v_s^2$ does not.

To give a concrete estimate for the size of $\mu$, we can simplify
the integral by considering the leading logarithmic behavior in the
gauge coupling constant $g$, which comes from $t$-channel exchange
at low momentum transfer, where ${\cal M}\sim 1/t$.  Using the same
approximations as in Appendix A of ref.~\cite{MP}, we obtain
\beq
	 \mu = {\zeta(3)\over 6\zeta(2)} v_s^2 T \cong 0.12 v_s^2 T,
	\qquad v_s\ll 1,
\label{smvres}
\eeq
in the small-velocity limit.  The value of $\mu$, relevant to
baryogenesis, is the sum of $1/2$ this quantity, over all left handed
doublets coupled to SU(2) which scatter from the wall in this way.

While we have only performed the calculation in the small $v_s$ case,
where it is valid to make an expansion about equilibrium population
functions, we expect that, for relativistic strings,
\beq
	\mu \lsim T,\qquad  \gamma_v v_s \sim 1 \, .
\label{lgvres}
\eeq
since parametrically there is nothing that can enhance $\mu$ to the
point of getting a highly degenerate gas of CP-asymmetric particles.
In fact, it will be seen that the arguments to be made below would not
be invalidated even if $\mu$ diverged like $\gamma_v$ in the
ultrarelativistic limit.

Eqs.~(\ref{smvres}-\ref{lgvres}) are sufficient for making our
estimate of the maximum string-mediated baryon asymmetry possible,
as will be done below in section 4.  To bolster and complete the
argument, however, we shall give some more realistic computations of
the CP asymmetry in the following subsections.

\subsection{Initial chiral flux in a realistic model}

We will now consider the generation of a CP-violating particle flux in
a specific theory.  The simplest example is a two-Higgs-doublet model
including~\cite{ckv} a complex mass term $\muu^2 \Phi_1^\dagger \Phi_2$
and quartic
interaction $h (\Phi_1^\dagger \Phi_2)^2$.  The unremovable
CP-violating phase $\theta_0$ is defined by $\muu^2 h^{1/2} = |\muu^2
h^{1/2}|e^{i\theta_0}$.  In the string wall, the two VEVs can be
parameterized as $\Phi_1(r) = \phi_1(r)e^{i\theta(r)/2}$ and $\Phi_2(r)
= \phi_2(r)e^{-i\theta(r)/2}$, where the $\phi_i$ functions are real
and $\delta\theta \equiv \theta(\infty)-\theta(0)$ is of order
$\theta_0$.  Solutions for $\theta(r)$ have been found for domain
(bubble) wall backgrounds in ref.~\cite{ckv}, where $\phi_i(r)$ has the
form $\tanh(r/r_0)$, and we expect the solutions to look similar in the
case of strings since the shape of our Higgs doublet profiles are
qualitatively the same.  We will therefore borrow results from
ref.~\cite{ckv} to make the following estimates.  This is done to be as
quantitative as possible, but our main conclusions will not depend on
fine details of this particular model, but rather on parametric
dependences that we expect to hold in any model.

If each fermion couples only to one of the two Higgs doublets, as is
beneficial for avoiding flavor-changing neutral currents, then they
will have CP-violating reflections from the string wall as a result of
the spatially varying phase $\theta(r)$.  Those particles with the
largest Yukawa couplings will be reflected most strongly.  For
simplicity we will model the string as having a square rather than
round cross section, moving through the plasma perpendicularly to one
of the sides ($\Box\to x$) with velocity $\vec v_s = v_s\hat x$.  While
this approximation is far from perfect, it can only change the answer
by a geometrical factor of order 1.  It allows the use of planar wall
calculations for the difference in reflection probabilities, which can
be adequately parameterized in terms of the fermion momentum
perpendicular to the string wall (which we will take to be $p_x$) by an
exponential,
\beq
   \Delta R(p_x) \cong A\, \delta\theta \, e^{-p_x/\delta_p}
\eeq
(the linear dependence on $\delta\theta$ being correct for
$\delta\theta\ll 1$), where $A$ and $\delta_p$ depend on the
mass $m_f$ of the fermion outside the string and the width of the Higgs 
field profile.  The dependence of $\delta_p$ was found to be
\beq
	{\delta_p\over m_f} =\left\{\begin{array}{ll} 
	-1.1\ln(m_f r_{1/2}) - 0.54, & m_f r_{1/2} < 0.3;\\ 
	0.19\, (m_f r_{1/2})^{-1.2}, & 0.3 < m_f r_{1/2} < 0.7;\\
	0.15\, (m_f r_{1/2})^{-1.8}, & m_f r_{1/2} > 0.7;
	\end{array}\right.
\eeq
It should be noticed that, although large values of $r_{1/2}$ are 
desirable for enhancing the rate of sphaleron interactions in the string,
they at the same time suppress the quantum reflections of heavy 
particles.  For this reason lighter particles like the $\tau$ lepton
can be more important than top quarks despite their smaller Yukawa 
couplings.

The amplitude of the asymmetry, $A$, depends not only on $m_f r_{1/2}$
but also on $\muu^2/m_h^2$.  For $m_f r_{1/2}\sim 1$, $A$ is also of order
$0.1$, but for larger $m_f r_{1/2}$ it decreases exponentially, $A \cong
e^{-c m r_{1/2}}$, where $c$ ranges from $0.5$ to $2$ as $\muu^2/m_h^2$
goes from 0.3 to 8.

Once $\Delta R$ is known, the next step is to compute the flux of CP
asymmetry going into the string.  
The flux of CP current per unit length of the string is given by
\beq
   J = {1\over 2}\int {d^2 p_\sperp dp_x\over (2\pi)^3} 
	\left|{p_x\over E}\right| \Delta R(p_x)\Bigl(
	f(\gamma_v(E+v_s \sqrt{p_x^2-m_f^2})/T) - 
	f(\gamma_v(E-v_s p_x)/T)\Bigr),
\eeq
where $f$ is the distribution function of the particles, boosted
by $\pm v$ (the wall velocity) depending on whether they are approaching
from the right or the left, and $\gamma_v = (1-v_s^2)^{-1/2}$.  Making
the Maxwell-Boltzmann approximation for the distribution functions,
and ignoring the dependence on $m_f$ (since the particles are
assumed to be relativistic) we can easily evaluate $J$ to be
\beq
	J = v_s \, A\, \delta\theta\,
	{ \delta_p^3\over \pi^2}\,{1+\gamma_v\delta_p/T\over
	\left( 1+2\gamma_v\delta_p/T + (\delta_p/T)^2
	\right)^2}
\eeq
In another theory the form of $J$ might differ, but it must always be
proportional to the string velocity, $v_s$, because at $v_s=0$
everything is in equilibrium.  Furthermore, it is bounded by $J \le v_s
T^3 / \pi^2$, which is the forward-backwards difference in total
incident particle numbers.  This bound would be saturated only if the
string were a perfect reflector of one CP state and a perfect
transmitter of the other; any realistic wall will give currents
that are much smaller.

\subsection{Subsequent diffusion of the source}

Now we are ready to use the current as a source for the diffusion
equation to find the steady-state density of chiral asymmetry at the
center of the string.  It must be understood that the flux $J$ 
actually represents a positive flux entering the front wall and going
into the string, and an equal and opposite flux leaving the
string through the front wall and entering the plasma.  To apply the
diffusion equation, it must be assumed that these fluxes penetrate the
plasma by some distance $\lambda$ (the mean free path) before diffusive
behavior sets in.  Otherwise the diffusion equation sees only two
equal and opposite, hence canceling, local densities at the string wall,
giving no net source of chiral charge.  As explained in references
\cite{ckv,jpt}, this can be modeled by a source term in the
diffusion equation which is a difference of two delta functions:
\beq
	{\partial n\over \partial t} - D \left(
	{\partial^2\over\partial x^2} + {\partial^2\over\partial y^2}
	\right) n +\Gamma n =  
	J \left(\delta(x+\lambda-v_s t)-\delta(x-\lambda-v_s t)\right)
\label{diffeq}
\eeq
Here $y$ is the transverse direction along the cross-section of the 
string, $D$ is the fermion's diffusion constant, and $\Gamma$ is the
rate of chirality-damping processes, like spin-flip interactions.

However, the above equation gives only the contribution from the front
wall.  There is another contribution from the back wall which has the
opposite sign.  Taking this into account, and changing
coordinates to the rest frame of the string, where $x=y=0$ denotes the
center, we find that the solution for $n$ can be expressed as a sum of
four contributions, two for each wall:
\beqa
	n(x,y) & = & J \Big[ N(x+w/2+\lambda,y) - N(x+w/2-\lambda,y)
	\nonumber\\
	&& \qquad  + N(x-w/2+\lambda,y) - N(x-w/2-\lambda,y)\Big] \, ,
\label{lincom}
\eeqa
where $w = 2r_{1/2}$ is the width of the string and $N$ is the Green's
function of the 2-D diffusion equation.  The first line represents the
front wall contribution and the second is that of the back wall.
Although the explicit signs in the equation are the same for both
walls, the contributions to the chemical potential at the center of the
string are of opposite sign; it is the second and third contributions
in the equation which are injected inside of the string.  It is only
because the string's motion leads to an asymmetric solution $N$ of the
diffusion equation that the expression for $n$ is nonzero at the center
of the string; in fact as the string velocity $v_s$ goes to zero $N$
becomes a symmetric function and the average over the string interior
of the term in brackets goes to zero linearly in $v_s$; this is whence
the second power of $v_s$ arises.

Using the Green's function for the two-dimensional diffusion equation, 
the individual contributions are given by 
\beqa
	N(x,y) = {1 \over 4\pi D} \int_0^\infty 
	{dx'\over x'}\int_{-w/2}^{w/2} dy'
	e^{-(v/4Dx')((x+x')^2 + (y-y')^2) - \Gamma x'/v},
\label{diffsoln}
\eeqa
which satisfies a two-dimensional diffusion equation like (\ref{diffeq})
with source term 
\beq
\delta(x)\int_{-w/2}^{w/2} dy'\delta(y-y').
\eeq
The derivation of eq.~(\ref{diffsoln}) is not completely obvious.  For
a stationary source, the Green's function for the 2-D diffusion
equation is $G(x,y,t) = t^{-1}\exp(-(x^2+y^2)/4Dt - \Gamma t)$.  Now we
are considering a moving source, whose trajectory can be described by
$x' = v t'$.  At a given position $(x,y)$ and time $t$, an observer
sees a contribution from when the source was at position $x'$ at the
earlier time $t'$,  proportional to $\int dy' G(x-x',y-y',t-t')$.  We
must integrate all these contributions from $t'=-\infty$ to the present
time $t$: $N(x,y,t) \propto \int_{-\infty}^t dt' \int dy'
G(x-x',y-y',t-t')$.  Here $N$ is still in the rest frame of the plasma;
replacing $x\to x+vt$ takes us to the rest frame of the source, so that
$x$ now measures how far one is in front of the source, and the
solution becomes time-independent. The change of variables $x' =
v(t-t')$ in the integral gives eq.~(\ref{diffsoln}).  As for the
normalization, it is chosen to insure that the rate of particle
creation by the source equals that of particle loss by decay when the
system reaches a steady state:
\beq
	\Gamma \int J\, N(x,y)\, dx\, dy = Jw.
\eeq

\begin{figure}[ht]
\centerline{\epsfxsize=6in\epsfbox{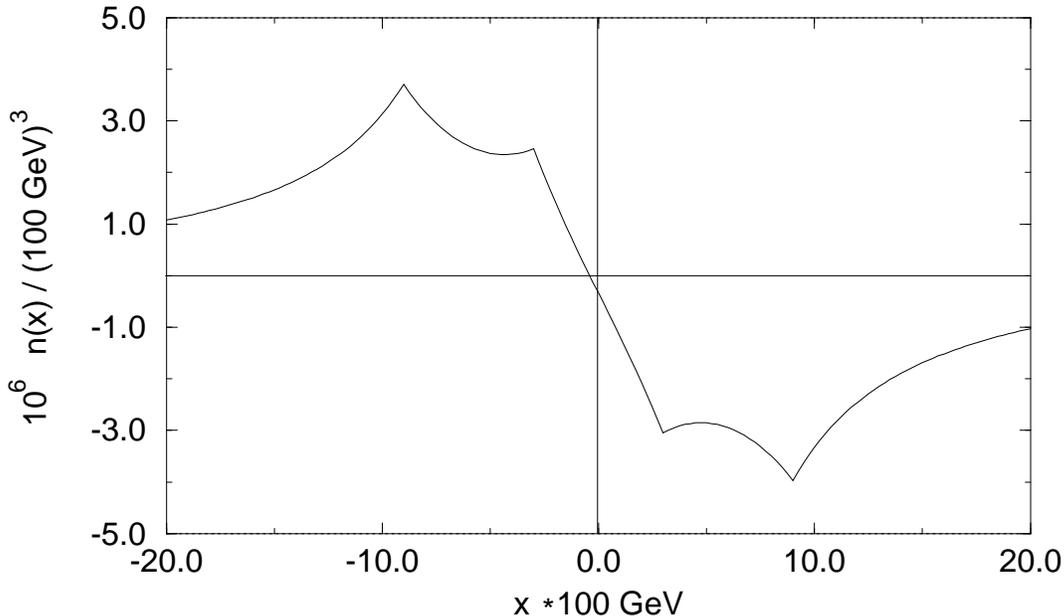}}
\caption{\label{denprofile} The chiral asymmetry as a function of 
distance from the center of the string, for the parameters
$D=6/T$, $\lambda = D$, $\Gamma = T/100$, $w=6/T$,
$\delta_p = 0.1 T$, and $v = 0.1$.}
\end{figure}

At the center of the string ($y=0$), where the effect is maximized,
the answer simplifies somewhat, and can be expressed as
\beq
	N(x)\equiv N(x,0) = {e^{-vx/2D} \over 4D\sqrt{\pi}}
   \int_0^\infty {dz\over\sqrt{z}}\, \erf(w /\sqrt{4z}) 
   e^{-x^2/z-(v/4D)(v/4D+\Gamma/v)z}.
\label{densoln}
\eeq
Substituting this result into eq.~(\ref{lincom}) and evaluating it
as a function of $x$, we find that the chiral density in the string
is almost antisymmetric about the core, but is skewed slightly by
the nonzero velocity, which distinguishes front from back.  This
behavior (and the fact that we have idealized the string by four
delta function sources, eq.~(\ref{lincom})) 
is shown in figure \ref{denprofile}.  Because $n(x)$ varies significantly
inside the core of the string, and the rate of baryon number violation
at position $x$ is proportional to $n(x)$, 
we should  average it over the core, 
{\it i.e.,} the region $x\in [-w/2,w/2]$:
\beq
	\bar n = {1\over w} \int_{-w/2}^{w/2} n(x,0)\, dx.
\label{avgn}
\eeq
We should also average over $y$, but $n$ is a smooth function of $y$,
peaking at $y=0$, so using $y=0$ only slightly overestimates the
average.  This approximation, in any case, errs in favor of the mechanism
we are ruling out.

\begin{figure}[ht]
\centerline{\epsfxsize=6in\epsfbox{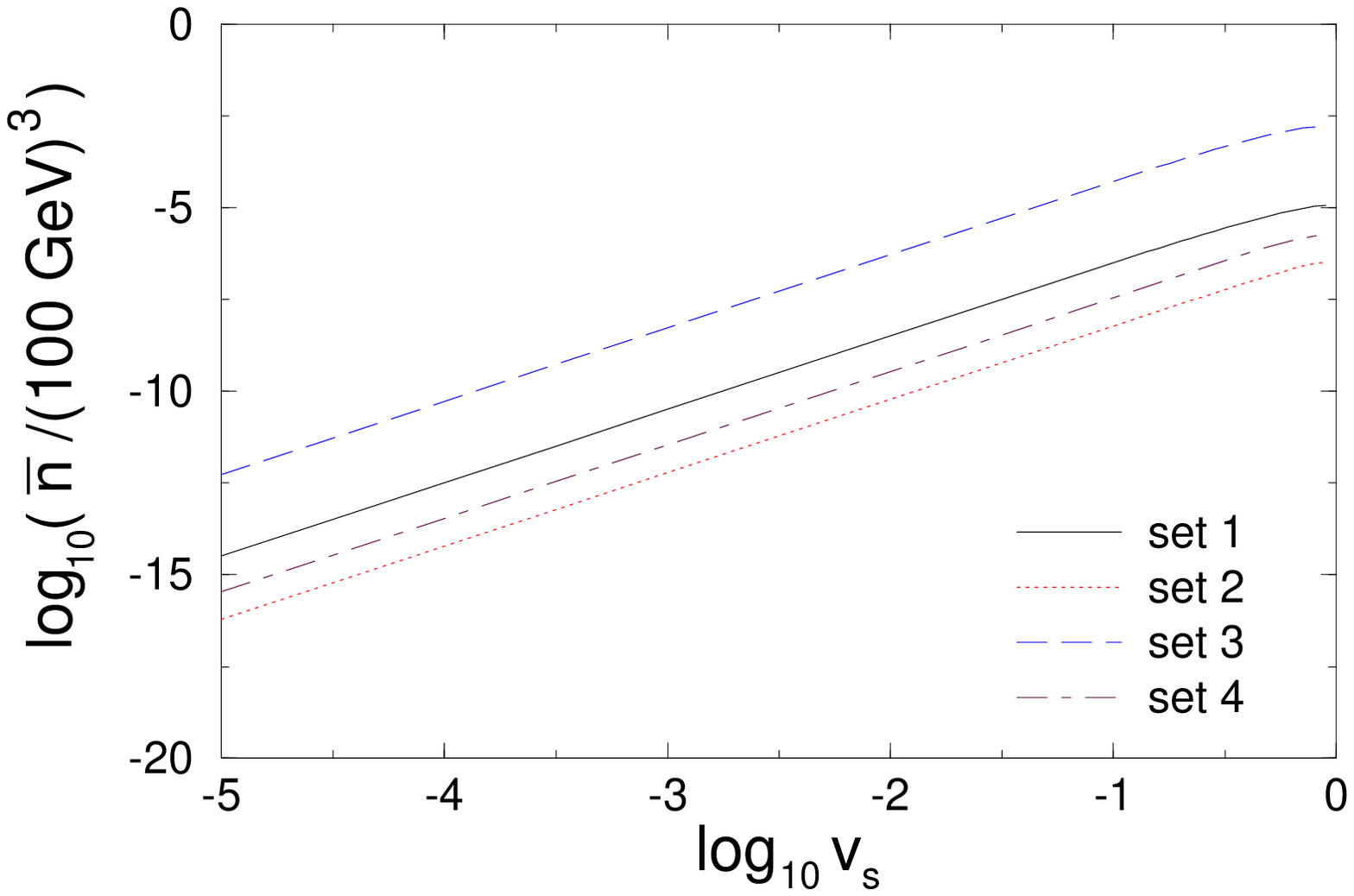}}
\caption{\label{lnnvsv} The chiral asymmetry averaged over the string
cross section, as a function of the string velocity, treating particle
reflections from the string quantum mechanically.  See text for
description of parameter sets.} 
\end{figure}

In Figure \ref{lnnvsv} we show how the spatially averaged 
chiral asymmetry at the center
of the string varies with the string velocity, for several choices of
parameters.  Set 1: $D=6/T$, $\lambda = D$, $\Gamma = T/100$, $w=6/T$,
$\delta_p = 0.1 T$, which is a realistic choice for quarks
\cite{jpt2}.  Set 2: same as set 1 but with $D=\lambda=100/T$, which
might be reasonable for leptons.  Set 3:  same as set 1 but with
$\delta_p = T$, to show what happens when the bubble surface
efficiently reflects particles with higher momenta. Set 4:  
same as set 1 but with $w=100/T$, to show
what happens for a very thick string.  In all examples we have assumed
that $T=100$ GeV, which is the temperature where the electroweak phase
transition is known to occur, and that CP violation is nearly maximal,
$\delta\theta = 1$.

In every case the number density, and hence the chemical potential
(related by $n = \mu T^2 /6$ times the number of spin degrees of
freedom), varies quadratically with the string velocity, $\mu/T = K
v_s^2$, with $K$ significantly less than $1$.  One power of $v_s$
arises from the injected flux, while a second power comes from a
front-back cancellation.  

To understand why the coefficient of the $v_s^2$ law is so small, it is
important to realize that the string core width is typically not much
larger than the quark diffusion constant, and so an asymmetry generated
on the string fairly rapidly diffuses away from it into the plasma at
large.  Since baryon number violation is only efficient right on the
string, any chiral asymmetry outside the core is wasted, as far as
baryogenesis is concerned, which helps to explain why the coefficient
of $\mu$ is so small. Even if we took the flux to be maximal the
coefficient $K$ would still be of order $10^{-2}$.  This is to be
contrasted with the situation in the conventional scenario with a first
order phase transition; here the wall is planar, so there is only one
direction to diffuse away from it, and baryon number is efficiently
violated in one of the half-spaces.

It should be mentioned that the $v_s$-dependence of the chemical potential
would be different for a very thick string.  When the string width satisfies
both $r > D / v_s$ and $r > T^3 / \Gamma_{\rm sph} \sim 10^6 / T$ (where
$\Gamma_{\rm sph}\sim 10^{-6}T^4$ is the rate per unit volume of sphaleron
interactions in the symmetric phase) then the baryon number asymmetry from the
front side vanishes; any baryons produced fall into the string and are
inevitably destroyed by sphalerons \cite{Tomislav}.  In this case the baryon
number production can scale linearly with $v_s$ (although at small enough
$v_s$, the $r > D / v_s$ condition will eventually break down).  However, the
nonsuperconducting strings we have considered have radii nowhere near this
large, and $\mu \propto v_s^2$ holds up to $v_s \sim 1$.

Our treatment becomes less good for $v_s$ approaching 1, but we do not 
expect a large CP asymmetry in this case either.
At such large velocities, the flux impinging on the back side of the
string can be ignored compared to that which hits the string's front
side and escapes through the back.  In the string rest frame, this
incident flux is composed of very hard particles.  The reflection
probability for hard particles drops very quickly with momentum, as was
mentioned above, and so therefore does the CP-violating
difference in reflection probabilities.  In any case, even in a scaling
string network where friction is negligible, 
almost none of the strings are traveling
ultrarelativistically; the mean value of $v_s^2$ (with the average
weighted by the energy and not the length of string) is less than 1/2
\cite{somestringguru}.  For a scaling-regime network what is relevant
is an appropriate average of $\mu(v_s)$ over the velocities represented
by the scaling network, for which we expect $\mu \sim T$ with a constant
of proportionality roughly of the same order as $K$ the coefficient in
the $v^2_s$ law discussed above.

\subsection{WKB estimate of the chiral asymmetry}
\begin{figure}[ht]
\centerline{\epsfxsize=6in\epsfbox{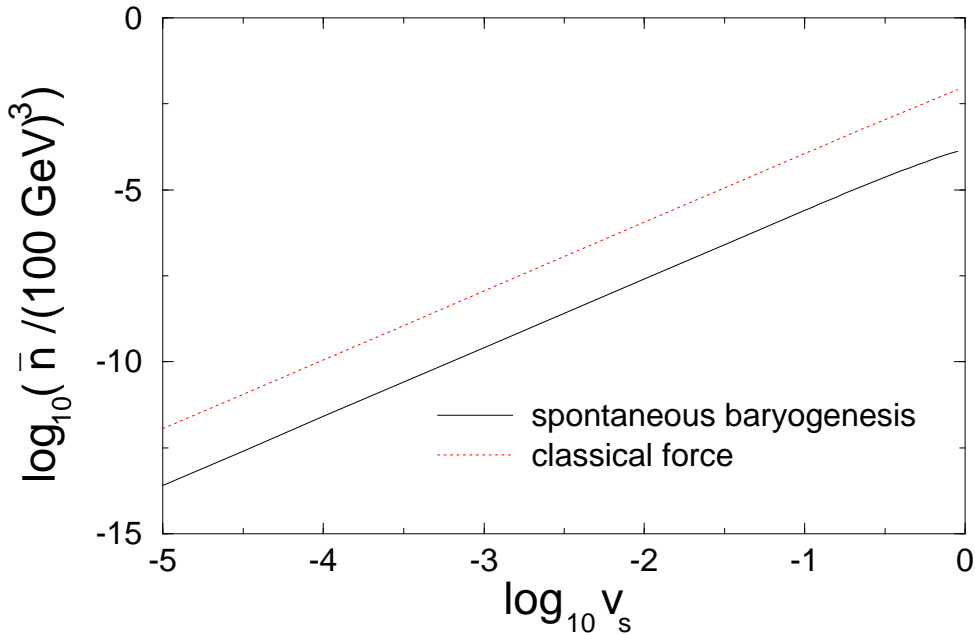}}
\caption{\label{lnnvsv_wkb} The chiral asymmetry at the center of the 
string as a function of $v_s$, using the WKB approximation,
and the parameters $D=6/T$, $\Gamma_{hf} = T/100$, $w=6/T$, 
$\delta\theta = 1$, $y_f=1$. }
\end{figure}

It has been pointed out that when defect boundaries are sufficiently thick
compared to the mean free path for particles to scatter in the plasma, a
semiclassical treatment of chiral asymmetry generation is more appropriate
than the quantum mechanical one given above \cite{jpt2,cjk}, and gives a
larger answer.  The distinction arises because CP violation in the defect
creates a classical, CP-violating force on particles as they cross the wall. 
In addition, the scatterings themselves can be CP-violating, giving rise to
what has been called ``spontaneous baryogenesis'' in earlier literature.  

If $\theta(x)$ is the background CP-violating phase, such as appears in the
two-Higgs-doublet model discussed above, then these two classical effects
generically give rise to a source term in the diffusion equation (replacing
the r.h.s.\ of eq.~(\ref{diffeq})) of the form
\beq
	S(x) \sim v_s \,{y_f^2 T^2\over 6}\,\left (D \theta''' 
	+ \Gamma_{hf} \theta'\right),
\eeq
where $y_f$ is the Yukawa coupling of the fermion to the Higgs field.
The primes denote spatial derivatives, and $\Gamma_{hf}$ is the rate of
helicity-flipping interactions.  The first term is due to the classical force,
and the second is the origin of spontaneous baryogenesis.  The size of the 
chemical potential thereby generated can be roughly estimated just as in 
the previous section if we approximate $\theta$ as a step function, so that
$\theta' \cong \delta\theta\, \delta(x)$.  Taking into account the 
separate contributions from the front and back walls of the string is
accomplished by replacing the above source term by
\beq
	S_2(x) = S(x+w/2) - S(x-w/2) 
\eeq
The minus sign occurs because the shape of the $\theta(x)$ profile on
the back wall is the mirror image of that on the front wall, and $S(x)$
involves an odd number of derivatives of $\theta(x)$.  This behavior is
in contrast to the quantum mechanical calculation, where front and back
walls gave same-sign contributions.  This difference between the
quantum and classical treatments stems from the fact that quantum
reflection probabilities are the same for particles crossing a
potential step, regardless of the direction of motion, whereas the
sign of the classical force exerted on the particle by such a step  
does depend on what direction it is going.

With the delta-function approximation for $\theta'$,
 we can immediately write the spatially averaged solution for the
chiral asymmetry in the string, in analogy to the quantum mechanical
treatment of the preceding section.  $\bar n$ is given by 
eq.~(\ref{avgn}), using 
\beq
	n(x) = v_s\,\delta\theta\,{y_f^2 T^2\over 6}\left. \left(
	D N''(x') + {\Gamma_{hf}} N(x') 
	\right)\right|^{x'=x+w/2}_{x'=x-w/2},
\eeq
and the definition of $N(x)$ given by eq.~(\ref{densoln}).  The
results for the previously-described parameter set 1 ($D=6/T$,
$\Gamma_{hf} = T/100$, $w=6/T$, $\delta\theta = 1$), assuming $y_f=1$,
are shown in figure \ref{lnnvsv_wkb}.  The two terms, due to the
spontaneous baryogenesis and classical force effects, respectively, are
shown separately.  The classical force contribution is larger than the
spontaneous baryogenesis contribution, and is also significantly larger
than the quantum mechanical prediction for the same set of parameters.
The $v_s^2$ dependence is evident over most of the range of
velocities.  To convert from density to chemical potential, one uses
\beq
	\mu = {6 n \over g T^2}
\eeq
for a fermion with $g$ helicity states.

\section{Efficiency of baryogenesis by a string network}

We have now seen how efficiently baryon number is violated along cosmic
strings.  Namely, the production of baryons per unit length of string is
\begin{equation}
\frac{dN_B}{dL dt} = 1.5 \kappa_l \alpha_W^2 T^2 
	\sum_i \frac{- \mu_i}{T} \, ,
\label{rateperL}
\end{equation}
where the $\mu_i$ are 
the chemical potentials for the $i$th species of left-handed (SU(2)
doublet) fermions. 
Henceforth we will write $\sum_i \mu_i$ simply as $\mu$.  We have already seen
that $\kappa_l < 0.1$, falling fairly rapidly as the temperature drops, and
that $\mu$ varies as $v_s^2$, the square of the string velocity, when $v_s$ is
small.  To estimate the baryon number produced, we still need to know the
density and velocity of the string network, however.  These quantities
depend crucially on one model-dependent parameter of the strings,
namely their tension.

Let us denote the string tension by $\tau$, the typical string's radius of
curvature  at the electroweak epoch by $R$, and the typical
velocity of propagation of a string, in the rest frame of the plasma, by $v$.
On dimensional grounds we have that
\begin{equation}
v \sim H R \, ,
\label{vHR}
\end{equation}
with $H$ the Hubble constant at the electroweak epoch.  

The string experiences a velocity-dependent friction as it moves through
the
plasma, arising from particles bouncing off the electroweak-symmetry-restored
core of the string.  For small velocities one can define a frictional constant
$\eta$, such that the force per unit length of string is $d\vec{F}/dL = - \eta
\vec{v}$.  If $w$ is the width of the symmetry-restored region, then
\beq
\eta \sim w \eta_{\rm b} \sim \frac{1}{g \phi_0} g^3 \phi_0^3 T \, ,
\eeq
where $\eta_{\rm b}$ is the friction constant of a moving electroweak bubble
wall, $\vec F / {\rm area} = - \vec{v} \eta_{\rm b}$, and its parametric
dependence, $\eta_{\rm b} \sim g^3 \phi_0^3 T$, was argued in
ref.~\cite{MooreTurok}.  If we take $\phi_0 \sim T$, as needed to
prevent baryon number violation in the bulk from erasing the baryons
produced by strings, then $\eta \sim g^2 T^3$; if instead we make the
parametric estimate appropriate right after a weak phase transition, 
$\phi_0 \sim g T$, we get $\eta \sim \alpha_w^2 T^3$.  This is the
friction on the string just from its dragging a 
region of restored electroweak symmetry through the plasma; there may also
be additional sources of friction, so the $\eta$ given above is a lower
bound. 

The force (per unit length) pulling a string forward is given by $\tau /
R$. If $v \ll 1$
then inertia is not important and we can equate the forces, and using
eq.~(\ref{vHR}), obtain the estimates
\begin{equation}
\frac{\tau}{R} = v \eta \quad \rightarrow \quad 
	v \sim \sqrt{ \frac{H \tau}{\eta}} \, , \,
	R \sim \sqrt{ \frac{\tau}{H \eta}} \, ,
\label{useful}
\end{equation}
valid whenever the resulting velocity is small.  This situation is called the
{\it friction-dominated regime}.  On the other hand, if the above estimate
gives $v \ge 1$, then friction is not important; instead the string evolution
will be dominated by the tension, expansion of the universe, and loss of
closed loops due to self-intersections.  In this {\it scaling regime}, $R \sim
1/H$ and $v \sim 1$, with a mean squared value $\langle v^2 \rangle < 0.5$
\cite{somestringguru}.  The crossover tension between the two cases for $T
\sim 100$ GeV is $\tau \ge ( 10^9 {\rm\ GeV})^2$. 

To obtain the rate of baryon number production per unit volume from
eq.~(\ref{rateperL}), we must calculate the total length of string per unit
volume.  One expects on dimensional grounds that the mean separation between
strings is $O(R)$, so the length of string per unit volume is $1/R^2$, and the
baryon density production rate is
\begin{equation}
\frac{dN_B}{dV dt} \sim \left( \frac{1}{R^2} \right) 
	\kappa_l \alpha_w^2 T^2 \frac{\mu}{T} \, .
\label{finalrate}
\end{equation}
As was discussed in the last section, at small velocities the chemical
potential behaves like $\mu = v^2 KT$, with $K$ a pure number depending on
$\phi_0 / T$ but not on $v$.  It was shown that $K$ cannot be larger than
$O(1)$, and realistically is much smaller.  Similarly, in the large-velocity
case, we expect $\mu \le T$, say, $\mu = K' T$. 

The two regimes of string evolution can now be considered in turn.
First consider the friction-dominated case.  Using $\mu = K v^2  T$ 
and eq.~(\ref{vHR}) in (\ref{finalrate}) gives
\begin{equation}
\frac{dN_B}{dV dt} \sim \kappa_l \alpha_w^2 K H^2 T^2 \, .
\end{equation}
This expression must be integrated from the time when the Higgs
condensate first becomes large enough to preserve baryon number in the
broken phase, $\phi_0 \simeq T$, to the present time.  The right hand
side depends on time through the temperature, $d(\ln T)/dt 
\sim H$.  The most strongly temperature-dependent term is $\kappa_l$,
which decreases quickly with falling temperature.  Even making the generous
assumption that it takes one Hubble time for $\kappa_l K$ to
fall sufficiently to cut off baryon number production, we still get a final
baryon density of order
\begin{equation}
\frac{N_B}{V T^3} \sim \kappa_l K \alpha_w^2 \frac{H}{T}
	\sim \kappa_l K \alpha_w^2 \frac{T}{m_{\rm pl}} \, .
\end{equation}
The last step follows from the Friedmann equation, $H^2 = (8 \pi / 3)  (\rho /
m_{\rm pl}^2)$, using $\rho \sim T^4$, so $H \sim T^2 / m_{\rm pl}$.  In this
expression, all dependence on the density of the string network, {\it i.e.,}
on $1/R$, has dropped out.  This happens because, although a denser network of
strings produces baryons along a greater total length of string, these strings
are moving more slowly and are therefore less efficient at baryon number
production. 

The above argument neglects the effects of small closed loops of string,
yet the string velocity in a friction-dominated network can
become large for a loop just as it collapses.  One may wonder whether such
loops can enhance the baryon production.  To answer this, 
consider a loop initially of radius $r=R$, the characteristic
curvature length of the network, which subsequently shrinks.  Its
velocity is $v(r) \propto 1/r$, and the radius evolves as $dr/dt = -v$.
Hence $r 
\propto (t_0-t)^{1/2}$, with $t_0$ the time of final collapse.  The baryon
production is proportional to
\begin{equation}
\int^{t_0} L v^2 dt = 2 \pi \int^{t_0} r v^2 dt 
	\propto \int^{t_0} (t_0-t)^{-1/2} dt \, ,
\end{equation}
which is well-behaved at the upper limit of integration and dominated by $r
\sim R$.  Hence the collapse of loops contributes approximately as much to the
baryon asymmetry as the evolution of the rest of the network, parametrically
no more.

Putting in the most optimistic values, $\kappa_l \sim 10^{-1}$, $K \sim
10^{-2}$, $T /m_{\rm pl} \sim 10^{-17}$, we find a baryon asymmetry about 13
orders of magnitude smaller than the physical value, $N_B / (V T^3) \sim N_B /
N_\gamma \sim 10^{-10}$.

If the string network is in the scaling regime, the previous argument changes
as follows.  Rather than $\mu = K v^2 T$, with $K \le 1$, we have $\mu = K' T$
with $K' \le 1$.  Also, $R \sim 1/H$ rather than $R \sim v/H$.  The final
expression is the same with $K'$ substituted for $K$, and the generated baryon
number is still too small by at least 13 orders of magnitude. 

These estimates are optimistic, in the sense that $\kappa_l$ falls quite
quickly with temperature, so there is much less than one Hubble time for
baryon number to be created; moreover $K$ is realistically much less than
$10^{-2}$.  The treatment of the defect network is somewhat pessimistic,
though, since numerical studies show that the mean separation of cosmic
strings is closer to $1/10$ of the Hubble length, rather than the full Hubble
length \cite{somestringguru}.  This could enhance our estimate of the baryon
asymmetry by a factor of $10^2$.  Also, the coefficients of order unity in the
Friedmann equation imply that $H \sim 10\, T^2 / m_{\rm pl}$, winning us an
additional factor of 10.  But this still leaves the mechanism too weak by 10
orders of magnitude, even using the most optimistic assumptions.

\medskip

{\large Interpretation:  Free Energy}

\medskip

There is another way of understanding our result, by considering the
available free energy in a network.  Think of the out of equilibrium
requirement of baryogenesis as the statement that a baryogenesis
mechanism must be a way of converting available free energy into baryon
number.  

The available free energy density in the conventional electroweak
baryogenesis scenario, with bubble walls converting supercooled
symmetric phase into broken phase, is the free energy difference between
the phases, which is $(T_c - T_{\rm nuc}) \Delta S$, the temperature
drop between equilibrium and nucleation times the entropy difference of
the phases.  The numerical value is parametrically $O(\alpha_W^{3} T^4)$
for $\lambda \sim g^2$, but it is always at least $10^{-3}T^4$ when the
phase transition is strong enough to prevent subsequent erasure of the
baryon number.

For the string network, the available free energy is the free energy
contained in the network, which is the length of string times the
tension (times the mean value of the relativistic $\gamma$ factor for
the scaling network, but this is order 1).  For the friction dominated
network, the density of the network is $\sim 1/R^2$, which together with 
Eq. (\ref{useful}) gives a free energy density $\sim H \eta \lsim 
T^4 (T/m_{\rm pl})$, independent of the tension.  Since the baryogenesis
mechanism is the same or similar as that in the conventional first order
case, it would be surprising if this much smaller available store of 
free energy were able to produce anything like the same number of
baryons.  For the scaling network the available free energy is larger,
though observational constraints require it to be 
less than $10^{-6} T^4$.  However, most of the free energy loss is not
due to friction against the plasma.  In fact, the amount which is being
lost to heating the plasma directly is independent of the tension of the
network, since the width of the symmetry restored region does not depend
on the string width.  In fact the free energy loss to the plasma is
about $\eta$ times the area swept out by the network in a Hubble time,
$\sim H \eta$, which is parametrically the same as for the friction
dominated network.  In either case, the network is just not capable of
generating anywhere near the amount of departure from equilibrium needed
to produce enough baryons.

The situation might be different for a superconducting network, but the
above reasoning suggests that even for this case, baryogenesis can only
get close to the efficiency of the conventional first order mechanism if
the free energy of the network is close to the observational limit {\em
and} much or most of the network's energy is lost to the plasma.

\section{Conclusion}

String networks do not provide a viable scenario for electroweak baryogenesis
if the strings are not superconducting.  While they are capable of generating
baryons, they do so far too inefficiently to account for the existing
abundance. 

The first problem is that baryon number violation along a string, while
certainly faster than in the broken phase, is not entirely unsuppressed.
The core where electroweak symmetry is restored is never wide enough to
allow symmetric phase sphaleron-like events; at best it brings the
sphaleron rate to $1/30$ of what it would be in the unsuppressed case,
{\it i.e.,} if the core region was $1/\alpha_w T$ in diameter.

Given the actual width of the region of symmetry restoration around a
cosmic string, we find that if the network is in the scaling
regime, it violates baryon number in far too small a total volume for
it to add up to the observed abundance.  If the network evolution is
friction-dominated, the network is denser, but the denser is the
network, the more slowly moving the strings are.  This in turn makes
them very inefficient at creating baryon number, since the production
rate goes as $v^2$ for small velocity.  In either case the generated
baryon number is at least 10 orders of magnitude smaller than the
observed abundance.

The case of superconducting cosmic strings merits more careful study.
The main challenge here is finding a realistic estimate of the
hypercharge current carried by a typical string at the electroweak
epoch.  We have not attempted to study this problem here.  However, our
results make it clear that unless the region of symmetry restoration
proves to be much wider than $1 / g^2 T$, the superconducting scenario
will also fail.

\section*{Acknowledgements}
We would like to thank R. Brandenburger, A. Kusenko, P. Langacker, and
M. Trodden for useful discussions.

\appendix

\section{Cooling to a saddle point}
\label{AppendixA}

Here we discuss an algorithm for numerically finding a saddle point of a
Hamiltonian on a many-dimensional space, which we used for determining
the energy for a sphaleron in a string background in Section \ref{Bviol}.

Suppose that we have a Hamiltonian system with degrees of freedom
$\Phi_\alpha$
and Hamiltonian $H(\Phi)$.  We require only that the coordinates
$\Phi$ are continuous.  They could for instance
be real numbers, or members of a compact manifold.

Ordinary gradient flow cooling of the fields means evolving the fields
under a cooling time $\tau$ according to
\begin{equation}
\frac{d\Phi_{\alpha}}{d\tau} = - \frac{\partial H(\Phi(\tau))}
	{\partial \Phi_\alpha} \, .
\end{equation}
When the $\Phi$ take values on a manifold the derivatives are understood
in terms of the tangent space at the current location.  A typical
discrete implementation of this algorithm is
\begin{equation}
\Phi_\alpha ( \tau + \Delta ) - \Phi_\alpha (\tau) =  - \Delta 
	\frac{ \partial H(\Phi(\tau))}{\partial \Phi_\alpha} \, .
\label{diss_algorithm}
\end{equation}
When $\Phi$ is a point on a manifold, the right hand side is a point in
the tangent space of $\Phi(\tau)$ and we determine $\Phi ( \tau +
\Delta)$ by starting at $\Phi(\tau)$ and following the geodesic curve of
starting direction and length indicated by the right hand side.

Provided that we make $\Delta$ small enough, this algorithm will reduce
the energy until it hits a local minimum.\footnote{It 
could in principle hit a
saddle point and stick there, but it would have to land exactly on it
and the space of starting points which would do so is measure zero.}  
``Small enough'' means
that $\Delta < 2 / \omega^2_{\rm max}$, where $\omega^2_{\max}$ is the
largest eigenvalue of the matrix of second derivatives of $H$.  In many
cases we can actually compute, or at least bound, $\omega^2_{\rm max}$
over the entire space,
and ensure stable dissipation.

What if we want an algorithm which will flow to the nearest extremum,
even if it is a saddle point and not a minimum?  If we can guess a
starting point which is fairly close to the desired extremum then it
makes sense to analyze algorithm behavior in terms of small deviations
about the extremum.  Write the fields as $\Phi_0 + \delta
\Phi_{\alpha}$, and expand the Hamiltonian to second order in $\delta
\Phi_{\alpha}$;
\begin{equation}
H \simeq H(\Phi_0) + H_{\alpha \beta} \delta \Phi_\alpha 
	\delta \Phi_\beta \, .
\end{equation}
The matrix $H_{\alpha \beta}$ is diagonalized by the basis of
eigenvectors $\xi_\alpha$, with eigenvalues $\lambda_\alpha$.  It is
convenient to write the distance from the extremum $\Phi_0$ in this
basis, as $\delta_\alpha \xi_\alpha$.  

The action of the algorithm, Eq. (\ref{diss_algorithm}), on
$\delta_\alpha$ is 
\begin{equation}
\delta_\alpha ( \tau + \Delta ) = [ 1 - \Delta \lambda_\alpha ] 
	\delta_\alpha ( \tau ) \, ,
\end{equation}
which shrinks those perturbations in directions with positive eigenvalue
and stretches perturbations in directions with negative eigenvalue.
This is fine if you want to find a minimum but it will carry you away
from a saddle point.

\begin{figure}[ht]
\centerline{\epsfxsize=4in\epsfbox{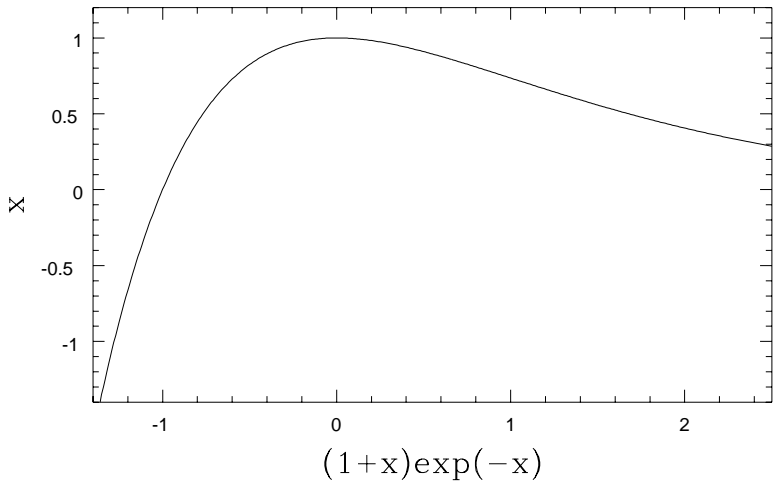}}
\caption{\label{simple} The function $(1+x)e^{-x}$.  The algorithm
converges for all eigenvalues $\lambda$ provided $N \Delta \lambda
\equiv x$ satisfies $-1 < (1+x) e^{-x} < 1$.}
\end{figure}

If on the other hand we take $N$ steps with step length $\Delta$ and one
step with step length $- N \Delta$, the departure from the saddle point
is 
\begin{equation}
\delta_\alpha ( {\rm after} ) = [ 1 + N \Delta \lambda_\alpha ] 
	[ 1 - \Delta \lambda_\alpha ]^N \delta_\alpha ( {\rm before} )
	\simeq (1 + N \Delta \lambda_\alpha) e^{- N \Delta
	\lambda_\alpha} \delta_\alpha({\rm before}) \, ,
\end{equation}
where the last approximation holds for $\Delta \lambda \ll 1$.
We plot the function $(1+x)e^{-x}$ in Figure \ref{simple}.  The
algorithm reduces the deviation in every basis direction provided that
the most negative eigenvalue $\lambda_{\rm neg}$ 
satisfies $1 + N \Delta \lambda_{\rm neg} e^{-N \Delta \lambda_{\rm neg}}
> -1$, which requires $N \Delta \lambda_{\rm neg} > -1.2784645$.  If we
choose $N \Delta$ too large, the deviation in the unstable direction
will alternate in sign and grow in magnitude; otherwise every departure
from the saddle point will shrink.

For our application we can tell if we have chosen too large a value for
$N \Delta$ by cooling the configuration towards vacuum and finding the
sign of ${\rm Tr} E \cdot B$ as the configuration slides out of the
saddle towards the vacuum.  If $N \Delta \lambda_{\rm neg} < -1$ then
the sign of the excitation in the unstable direction switches each step,
and so will the direction of approach towards the vacuum.  If the sign
switches we may be driving the system dangerously.  If it does not, we
are being too conservative and the algorithm will be inefficient.  We
can adjust $N \Delta$ as we go to achieve optimal stable performance.
If we did not have such a clear signature for the sign of the unstable
mode, we could still determine $\lambda_{\rm neg}$, once we get close to
the saddle point, by tracking the energy during cooling from near the
saddle to the vacuum.  After transients from stable modes have decayed,
but before the configuration gets far from the saddle, the energy
behaves like $E_{\rm sph} - K \exp(\tau \lambda_{\rm neg})$, and
$\lambda_{\rm neg}$ can be determined by fitting.

Also note that, if we know $\lambda_{\rm neg}$, and also know that there
is only one unstable mode, we can increase the efficiency with which the
algorithm eliminates low frequency stable modes by applying extra
cooling.  For instance we can perform a backwards step of length $1 /
\lambda_{\rm neg}$ and several small forward steps of total length $2 /
| \lambda_{\rm neg}|$, so the departure from the saddle point obeys
\begin{equation}
\delta_{\alpha}({\rm after}) = ( 1 + \lambda_\alpha / \lambda_{\rm neg}) 
	\exp(- 2 \lambda_\alpha / \lambda_{\rm neg}) \delta_{\alpha}
	( {\rm before} ) \, .
\end{equation}
However, if we expect there to be more than one negative eigenmode then
this is may make the other one grow unstably.  In general, if there are
more than 1 negative eigenmodes and there are low lying positive
eigenmodes, the algorithm becomes quite inefficient.  Also note that the
algorithm  will not necessarily find a saddle point if 
it starts out very far from it.  Both of these limitations are quite general
limitations of saddle point finding algorithms.  The algorithm's charm is
its simplicity; all we need to be able to do is take first derivatives
of the Hamiltonian in order to apply it, though we should try to
determine $\lambda_{\rm neg}$ if we want to apply it well.  
No more coding is necessary than it takes already to search for minima.

\end{document}